\documentclass[%
 reprint,
superscriptaddress,
nofootinbib,
 amsmath,amssymb,
 aps,
]{revtex4-1}

\usepackage{graphicx}
\usepackage{dcolumn}
\usepackage{bm}
\usepackage{xcolor}

\usepackage{subfigure}

\begin{document}

\preprint{APS/123-QED}

\title{Hidden Markov model tracking of continuous gravitational waves \\from young supernova remnants}

\author{L. Sun}
\email{lings2@student.unimelb.edu.au}
\author{A. Melatos}
\email{amelatos@unimelb.edu.au}
\affiliation{\mbox{ozGrav, University of Melbourne, Parkville, Victoria 3010, Australia}}
\author{S. Suvorova}
\affiliation{\mbox{School of Electrical and Computer Engineering, RMIT University, Melbourne, Victoria 3000, Australia}}
\affiliation{\mbox{ozGrav, University of Melbourne, Parkville, Victoria 3010, Australia}}
\author{W. Moran}
\affiliation{\mbox{School of Electrical and Computer Engineering, RMIT University, Melbourne, Victoria 3000, Australia}}
\author{R. J. Evans}
\affiliation{\mbox{Department of Electrical and Electronic Engineering, University of Melbourne, Parkville, Victoria 3010, Australia}}
\affiliation{\mbox{ozGrav, University of Melbourne, Parkville, Victoria 3010, Australia}}
\date{\today}

\begin{abstract}

Searches for persistent gravitational radiation from nonpulsating neutron stars in young supernova remnants (SNRs) are computationally challenging because of rapid stellar braking. We describe a practical, efficient, semi-coherent search based on a hidden Markov model (HMM) tracking scheme, solved by the Viterbi algorithm, combined with a maximum likelihood matched filter, the $\mathcal{F}$-statistic. The scheme is well suited to analyzing data from advanced detectors like the Advanced Laser Interferometer Gravitational Wave Observatory (Advanced LIGO). It can track rapid phase evolution from secular stellar braking and stochastic timing noise torques simultaneously without searching second- and higher-order derivatives of the signal frequency, providing an economical alternative to stack-slide-based semi-coherent algorithms. One implementation tracks the signal frequency alone. A second implementation tracks the signal frequency and its first time derivative. It improves the sensitivity by a factor of a few upon the first implementation, but the cost increases by two to three orders of magnitude.

\begin{description}
\item[PACS numbers]
95.85.Sz, 97.60.Jd
\end{description}
\end{abstract}

\pacs{Valid PACS appear here}
\maketitle


\section{Introduction}
Rotating neutron stars in young supernova remnants (SNRs) are plausible sources of quasi-monochromatic gravitational radiation detectable by ground-based interferometers such as the Laser Interferometer Gravitational Wave Observatory (LIGO) and the Virgo detector \cite{Abbott2007S2,2012-S6,Riles2013}.  The emission is predicted to occur at a frequency proportional to the star's spin frequency $f_\star$. A thermoelastic \cite{ushomirsky00, Johnson-McDaniel2013} or magnetic \cite{Cutler2002, Mastrano2011, Lasky2013} mass quadrupole emits at $f_\star$ and/or $2f_\star$, an r-mode current quadrupole emits at roughly $4{f_\star}/3$, perturbed by an equation-of-state correction \cite{Owen1998, Heyl2002, Arras2002, bondarescu09}, and a current quadrupole produced by nonaxisymmetric circulation in a superfluid pinned to the crust emits at $f_\star$ \cite{Peralta2006, VanEysden2008, Bennett2010, Melatos2015}. There are several reasons to devote attention to this source class. First, a young object has spent less time settling down since its birth; slow, diffusive processes like ohmic \cite{Haensel1990} or thermal \cite{Gnedin2000} relaxation are still in the early stages of erasing nonaxisymmetries inherited at birth \cite{Abbott2007S2,Knispel2008,Riles2013}. Second, the rapid spin down of a differentially rotating young neutron star excites high-Reynolds-number turbulence, which produces a time-varying current quadrupole moment \cite{Peralta2006,Melatos2010,Melatos2012}. Third, young radio pulsars are known to undergo glitches \cite{McKenna1990,Shemar1996,Urama1999,Melatos2008}, which are ascribed to differential rotation \cite{Mastrano2005,MelatosPeralta2007,Glampedakis2009} or starquakes \cite{Middleditch2006} and can also lead to quadrupole moment variations.

Initial LIGO achieved its design sensitivity over a wide frequency band during Science Run 5 (S5) \cite{lscinstrument09} and exceeded it during Science Run 6 (S6) \cite{2012-S6}. Several continuous-wave searches targeting young SNRs have been carried out using Initial LIGO data. A directed, radiometer search for SNR 1987A was conducted in S5, yielding a 90\% confidence strain upper limit $h_0^{90\%} = 7 \times 10^{-25}$ for a circularly polarized signal in the most sensitive band, near 160 Hz \cite{Eric2011}\footnote{The radiometer search assumes a circularly polarized signal. The upper limit quoted here converts to $h_0^{90\%} = 1.6 \times 10^{-24}$ for the general case of arbitrary polarization after multiplying by a sky-position-dependent factor $\approx 2.2$ \cite{MessengerNote}.}. A semi-coherent cross-correlation search for the same target in S5 data improved the upper limit to $h_0^{90\%} = 3.8 \times 10^{-25}$ \cite{Sun2016}. A coherent search for Cassiopeia A (Cas A) was conducted on a 12-day stretch of S5 data in the band 100--300\,Hz, yielding $h_0^{95\%}$ in the range 0.7--$1.2\times 10^{-24}$ \cite{Abadie2010}. The upper limit for Cas A was improved by approximately a factor of two by a semi-coherent Einstein@Home search in S6. The search was conducted in a broad frequency band 50--1000\,Hz, yielding the best $h_0^{90\%} \approx 2.9 \times 10^{-25}$ at 170\,Hz \cite{EH_CasA_S6}. In S6, directed searches were conducted for nine nonpulsating X-ray sources (central compact objects) in SNRs with the maximum likelihood matched filter called the $\mathcal{F}$-statistic. The searches combine multi-detector data coherently over 5.3 to 25.3 days, yielding strain upper limits in the range $3.7 \times 10^{-25} \leq h_0^{95\%} \leq 6.4 \times 10^{-25}$ \cite{Aasi2015-snr}. The foregoing upper limits are slated to improve in the future. The strain noise in the first observation run (O1) of Advanced LIGO is 3--4 times lower than in S6 across the most sensitive band, between 100\,Hz and 300\,Hz, and $\sim 10^2$ times lower around 50\,Hz \cite{Abbott2016-detector}. The sensitivity is expected to improve roughly two-fold relative to O1 after further upgrades \cite{Abbott2016-detector}. 

The rapid spin down of young neutron stars is a serious challenge for searches of the above kind. A large number of matched filters (i.e. templates) is required to track a signal with rapid evolving phase, when a radio ephemeris is unavailable. For example, the $\mathcal{F}$-statistic search in Ref. \cite{Aasi2015-snr} is restricted to $\leq 25.3$\,d in order to keep the number of matched filters manageable. Semi-coherent methods have been developed \cite{Stackslide,PowerFlux01,dhurandhar08}, based on the stack-slide algorithm \cite{Brady2000}, to sum the signal power in multiple coherent segments after sliding the segments in the frequency domain to account for the phase evolution of the source. However, stack-slide searches are still computationally challenging, when high-order derivatives of the signal frequency enter the phase model. Intrinsic, stochastic $f_\star$ wandering (`timing noise') also degrades the sensitivity of these searches.

In this paper, we introduce an approach based on a hidden Markov model (HMM) \cite{Quinn2001} to tackle the above challenge. A HMM tracks signals with time-varying, unobservable parameters by modeling them as hidden states in a Markov chain. The HMM relates the observed data to the hidden states through a likelihood statistic and infers the most probable sequence of hidden states. It incoherently combines the coherent matched filter outputs from data blocks of duration $T_{\rm drift}$ (analogous to $T_{\rm span}$ in Ref. \cite{Aasi2015-snr}), during which the signal parameters are assumed to remain constant. The sensitivity scales approximately proportional to $\sim (T_{\rm obs}T_{\rm drift})^{-1/4}$, where $T_{\rm obs}$ is the whole observation time. The Viterbi algorithm \cite{Viterbi1967} provides a computationally efficient HMM solution. A HMM was applied to search for continuous gravitational radiation from the most luminous low-mass X-ray binary, Scorpius X-1, in O1 data, taking into account the effects of spin wandering caused by the fluctuating accretion torque \cite{ScoX1Viterbi2017}.

The structure of the paper is as follows. In Section \ref{sec:matched_filter}, we describe the signal model, introduce the $\mathcal{F}$-statistic, and discuss the search parameter ranges. In Section \ref{sec:hmm_f0}, we formulate the HMM tracking problem with one hidden state variable, describe how to choose $T_{\rm drift}$, and discuss the impact of timing noise. In Section \ref{sec:simulations_and_sensi}, we conduct Monte-Carlo simulations in Gaussian noise, present search examples, and estimate the sensitivity. In Section \ref{sec:hmm_fdot}, we introduce an alternative HMM formulation with two hidden state variables, and present abridged simulation examples. In Section \ref{sec:discussion}, we discuss the trade-off between computing cost and sensitivity. We also discuss the special case of young objects, whose current spin frequencies are close to the value at birth. A summary of the conclusions is provided in Section \ref{sec:conclusion}.

\section{Coherent matched filter}
\label{sec:matched_filter}
 In this section we start by describing the signal model in Section \ref{sec:sig_model}. We then review the maximum likelihood matched filter corresponding to the signal model, called the $\mathcal{F}$-statistic, in Section \ref{sec:F-stat}, and discuss the signal phase parameter ranges in Section \ref{sec:search_para_range}.

\subsection{Signal model}
\label{sec:sig_model}
We consider a continuous gravitational wave signal from a rotating neutron star modeled as a biaxial rotor. The Doppler modulation of the observed signal frequency due to the motion of the Earth with respect to the solar system barycentre (SSB) is taken into consideration. The signal phase observed at the detector is then given by \cite{Jaranowski1998}
\begin{equation}
	\label{eqn:phase}
	\Phi(t) = 2\pi \sum_{k=0}^{s} \frac{f_0^{(k)}t^{k+1}}{(k+1)!}+ \frac{2\pi}{c}\hat{n}\cdot\vec{r}(t)\sum_{k=0}^{s}\frac{f_0^{(k)}t^k}{k!},
\end{equation}
where $f_0^{(k)}$ is the $k$-th time derivative of the signal frequency at $t=0$, $\hat{n}$ is the unit vector pointing from the SSB to the neutron star, and $\vec{r}(t)$ is the position vector of the detector relative to the SSB. 

The signal can be written in the form
\begin{equation}
\label{eqn:sig_form}
h(t)=\mathcal{A}^\mu h_\mu(t),
\end{equation}
where $\mathcal{A}^\mu$ denotes the amplitudes associated with the four linearly independent components\footnote{Here we assume a perpendicular rotor emitting at $2 f_\star$ only for simplicity. A non-perpendicular rotor also emits at $f_\star$, and hence eight components are involved. A full description of the signal model can be found in Ref. \cite{Jaranowski1998}. The emission spectrum of a triaxial rotor contains additional lines \cite{Lasky2013,VanDenBroeck2005}.}
\begin{eqnarray}
	\label{eqn:h1} h_1(t) &=& a(t) \cos \Phi(t), \\
	h_2(t) &=& b(t) \cos \Phi(t), \\
	h_3(t) &=& a(t) \sin \Phi(t),  \\
	\label{eqn:h4} h_4(t) &=& b(t) \sin \Phi(t),
\end{eqnarray}
$a(t)$ and $b(t)$ are the antenna-pattern functions defined by Equations (12) and (13) in Ref. \cite{Jaranowski1998}, and $\Phi(t)$ is the signal phase given by (\ref{eqn:phase}). In (\ref{eqn:sig_form}), $\mathcal{A}^\mu$ depends on the star's inclination, wave polarization, initial phase at $t=0$ and strain amplitude $h_0$.

\subsection{$\mathcal{F}$-statistic}
\label{sec:F-stat}
The time-domain data collected by a detector takes the form
\begin{equation}
x(t)=\mathcal{A}^\mu h_\mu(t) + n(t),
\end{equation}
where $n(t)$ stands for stationary, additive noise. The $\mathcal{F}$-statistic maximizes the likelihood of detecting a signal in data $x(t)$ with respect to $\mathcal{A}^\mu$ \cite{Jaranowski1998}. We define a scalar product $(\cdot|\cdot)$ as a sum over single-detector inner products,
\begin{eqnarray}
(x|y) &=& \mathop{\sum} \limits_{X} (x^X|y^X) \\
&=& \mathop{\sum} \limits_{X} 4\Re \int_{0}^{\infty}df \frac{\tilde{x}^X(f)\tilde{y}^{X*}(f)}{S_h^X(f)},
\end{eqnarray}
where $X$ indexes the detector, $S_h^X(f)$ is the single-sided power spectral density (PSD) of detector $X$, the tilde denotes a Fourier transform, and $\Re$ returns the real part of a complex number \cite{Prix2007}. The $\mathcal{F}$-statistic is expressed in the form
\begin{equation}
\mathcal{F} = \frac{1}{2} x_\mu \mathcal{M}^{\mu \nu} x_\nu,
\end{equation}
where we write $x_\mu = (x|h_\mu)$, and $\mathcal{M}^{\mu \nu}$ denotes the matrix inverse of $\mathcal{M}_{\mu \nu}=(h_\mu|h_\nu)$.  

Assuming the noise $n(t)$ is Gaussian, the random variable $2\mathcal{F}$ follows a non-central chi-squared distribution with four degrees of freedom, whose probability density function (PDF) is
\begin{equation}
\label{eqn:chi2-dist}
p(2\mathcal{F})=\chi^2(2\mathcal{F}; 4,\rho_0^2), 
\end{equation}
with non-centrality parameter \cite{Jaranowski1998}
\begin{equation}
\rho_0^2 = \mathcal{A}^\mu \mathcal{M}_{\mu \nu} \mathcal{A}^\nu.
\end{equation}
Without a signal, the PDF of $2\mathcal{F}$ centralizes to $p(2\mathcal{F})=\chi^2(2\mathcal{F}; 4,0)$. Given a signal in Gaussian noise and assuming the same single-sided PSD, $S_h(f)$, in all detectors, the optimal signal-to-noise ratio equals $\rho_0$, given by \cite{Jaranowski1998,Suvorova2016}
\begin{equation}
\label{eqn:snr2}
\rho_0^2=\frac{K h_0^2T_\text{drift}}{S_h(f)},
\end{equation}
where the constant $K$ depends on the sky location, orientation of the source and the number of detectors, and $h_0$ denotes the characteristic gravitational-wave strain.

We leverage the existing, fully tested $\mathcal{F}$-statistic software infrastructure in the LSC Algorithm Library Applications (LALApps)\footnote{http://software.ligo.org/docs/lalsuite/lalapps/index.html} to compute $\mathcal{F}$ as a function of frequency and its time derivatives over an interval of length $T_{\rm drift}$ \cite{F-stat2011}. The software operates on the raw data collected by LIGO in the form of short Fourier transforms (SFTs), usually with length $T_{\rm SFT}=30$\,min for each SFT. It provides options to search up to the third time derivative of frequency, $\dddot{f_0}$. The implementations described in Section \ref{sec:hmm_f0} and \ref{sec:hmm_fdot} use the options to search over $f_0$ and ($f_0,\dot{f_0}$), respectively.

\subsection{Search parameter ranges}
\label{sec:search_para_range}
The ranges of $\dot{f_0}$ and $\ddot{f_0}$ to be considered in defining the parameters of the search can be reexpressed in terms of the range of braking index $n=f_0\ddot{f_0}/\dot{f_0}^2$ and the spin-down age of the source $\tau$, given by \cite{Abadie2010,Aasi2015-snr}
\begin{equation}
\label{eqn:dotf_range}
-\frac{f_0}{(n_{\rm min}-1)\tau}\leq \dot{f_0} \leq - \frac{f_0}{(n_{\rm max}-1)\tau},
\end{equation}
and
\begin{equation}
\label{eqn:ddotf_range}
\frac{n_{\rm min}\dot{f_0}^2}{f_0} \leq \ddot{f_0} \leq \frac{n_{\rm max}\dot{f_0}^2}{f_0}.
\end{equation}
Radio timing observations yield $1.4\leq n \leq 3$ for all pulsars, where $\ddot{f_0}$ can be measured reliably by absolute pulse numbering \cite{Owen1998,Wette2008} (cf. \cite{Andersson2017}). Gravitational radiation in the mass and current quadrupole channels corresponds to $n=5$ and $n=7$, respectively. In this quick study, we assume $2\leq n \leq 7$. Strictly speaking, the ranges obtained from (\ref{eqn:dotf_range}) and (\ref{eqn:ddotf_range}) are wider than needed. In a real search, one should ideally calculate the $\dot{f_0}$ and $\ddot{f_0}$ ranges for each fixed $2\leq n \leq 7$ and choose the widest $\dot{f_0}$ and $\ddot{f_0}$ ranges. Equations (\ref{eqn:dotf_range}) and (\ref{eqn:ddotf_range}) are valid, provided that $f_0{(t)}$ during the observation is much smaller than $f_0$ at birth, ${f_0}_{\rm birth}$. The regime $f_0(t)\sim {f_0}_{\rm birth}$ is considered in Section \ref{sec:spindown_vs_age}.

\section{HMM Tracking of $f_0$}
\label{sec:hmm_f0}

We begin this section by reviewing briefly the use of HMM tracking in gravitational wave searches (Section \ref{sec:hmm}). We formulate the tracker as a one-dimensional HMM with a single hidden variable $f_0$ (Section \ref{sec:f_tracking}), discuss the coherent drift time-scale $T_{\rm drift}$ (Section \ref{sec:T_drift}), and illustrate that the method can track secular spin down and stochastic timing noise simultaneously (Section \ref{sec:timning noise}). A full description of HMMs can be found in Ref. \cite{Suvorova2016}. The Viterbi algorithm used for solving the HMM is described in Appendix \ref{sec:viterbi}. 

\subsection{HMM formulation}
\label{sec:hmm}
A Markov chain is a stochastic process transitioning between discrete states at discrete times $\{t_0, \cdots, t_{N_T}\}$. A HMM is an automaton, in which the state variable $q(t) \in \{q_1, \cdots, q_{N_Q}\}$ is hidden (unobservable), and the measurement variable $o(t)\in \{o_1, \cdots, o_{N_O}\}$ is observable. The hidden state at time $t_{n+1}$ only depends on the state at time $t_n$ with transition probability
\begin{equation}
	\label{eqn:prob_matrix}
	A_{q_j q_i} = P [q(t_{n+1})=q_j|q(t_n)=q_i].
\end{equation}
The hidden state $q_i$ is observed in state $o_j$ with emission probability
\begin{equation}
	\label{eqn:likelihood}
	L_{o_j q_i} = P [o(t_n)=o_j|q(t_n)=q_i].
\end{equation}
Given the prior defined by
\begin{equation}
	\Pi_{q_i} = P [q(t_0)=q_i],
\end{equation}
the probability that the hidden state path $Q=\{q(t_0), \cdots, q(t_{N_T})\}$ gives rise to the observed sequence $O=\{o(t_0), \cdots, o(t_{N_T})\}$ via a Markov chain equals
\begin{equation}
	\label{eqn:prob}
	\begin{split}
		P(Q|O) = & L_{o(t_{N_T})q(t_{N_T})} A_{q(t_{N_T})q(t_{N_T-1})} \cdots L_{o(t_1)q(t_1)} \\ 
		& \times A_{q(t_1)q(t_0)} \Pi_{q(t_0)}.
	\end{split}
\end{equation}
The most probable path maximizing $P(Q|O)$, viz.
\begin{equation}
	Q^*(O)= \arg\max P(Q|O),
\end{equation}
gives the best estimate of $q(t)$ over the total observation, where $\arg \max (\cdots)$ returns the argument that maximizes the function $(\cdots)$.

\subsection{Transition and emission probabilities}
\label{sec:f_tracking}

We consider the one-dimensional hidden state variable $q(t)=f_0(t)$. The discrete hidden states are mapped one-to-one to the frequency bins in the output of a frequency-domain estimator $\mathcal{F}(f_0)$ computed over an interval of length $T_{\rm drift}$, with bin size $\Delta f_0$ selected using the metric described in Appendix \ref{sec:cost}. For simplicity here, we take $\Delta f_0 = 1/(2 T_{\rm drift})$, with mismatch $m \leq 0.2$. The mismatch $m$ is defined as the fractional reduction of $\mathcal{F}$-statistic power caused by discrete parameter sampling (see Appendix \ref{sec:cost}). We choose $T_{\rm drift}$ as described in Section \ref{sec:T_drift} to satisfy
\begin{equation}
\label{eqn:int_T_drift}
\left|\int_t^{t+T_{\rm drift}}dt' \dot{f_0}(t')\right| < \Delta f_0
\end{equation}
for $0<t<T_{\rm obs}$. 

In this section, we firstly consider the situation where the time-scale of timing noise, which causes $f_0$ to walk randomly, is much longer than the spin-down time-scale. Hence the impact of timing noise is negligible compared to secular spin down. Modifications needed to tackle stronger timing noise are discussed in Section \ref{sec:timning noise}. If we substitute the maximum $|\dot{f_0}|$ from (\ref{eqn:dotf_range}), denoted by $|\dot{f_0}|_{\rm max}$, into (\ref{eqn:int_T_drift}) and assume that $\dot{f_0}$ is uniformly distributed in the range from zero to $|\dot{f_0}|_{\rm max}$,\footnote{According to equation (\ref{eqn:dotf_range}), we have $|\dot{f_0}|_{\rm max} = 6|\dot{f_0}|_{\rm min}$, where $|\dot{f_0}|_{\rm min}$ is the minimum $|\dot{f_0}|$ from (\ref{eqn:dotf_range}). In practice, however, we normally search $|\dot{f_0}|$ values more than an order of magnitude smaller than $|\dot{f_0}|_{\rm max}$. Hence the search range of $\dot{f_0}$ is dominated by $|\dot{f_0}|_{\rm max}$, and we approximate the search range to be $0 \leq |\dot{f_0}| \leq |\dot{f_0}|_{\rm max}$.} equation (\ref{eqn:prob_matrix}) simplifies to
\begin{equation}
\label{eqn:trans_matrix}
A_{q_{i-1} q_i} = A_{q_i q_i} = \frac{1}{2},
\end{equation}
with all other entries being zero. By the definition of the frequency domain estimator, the emission probability is given by \cite{Jaranowski1998,F-stat2011,Suvorova2016}
\begin{eqnarray}
\label{eqn:emi_prob_matrix}
L_{o(t) q_i} &=& P [o(t)|{f_0}_i \leq f_0(t) \leq {f_0}_i+\Delta f_0]\\ 
\label{eqn:matrix_propto}
&\propto& \exp[\mathcal{F}({f_0}_i)],
\end{eqnarray}
during the interval [$t,t+T_{\rm drift}$], where ${f_0}_i$ is the value of $f_0$ in the $i$-th $f_0$ bin. Since we have no independent knowledge of $f_0$, we choose a uniform prior, viz.
\begin{equation}
\Pi_{q_i} = N_Q^{-1}.
\end{equation}

\subsection{Drift time-scale}
\label{sec:T_drift}
\begin{figure*}
	\centering
	\scalebox{0.48}{\includegraphics{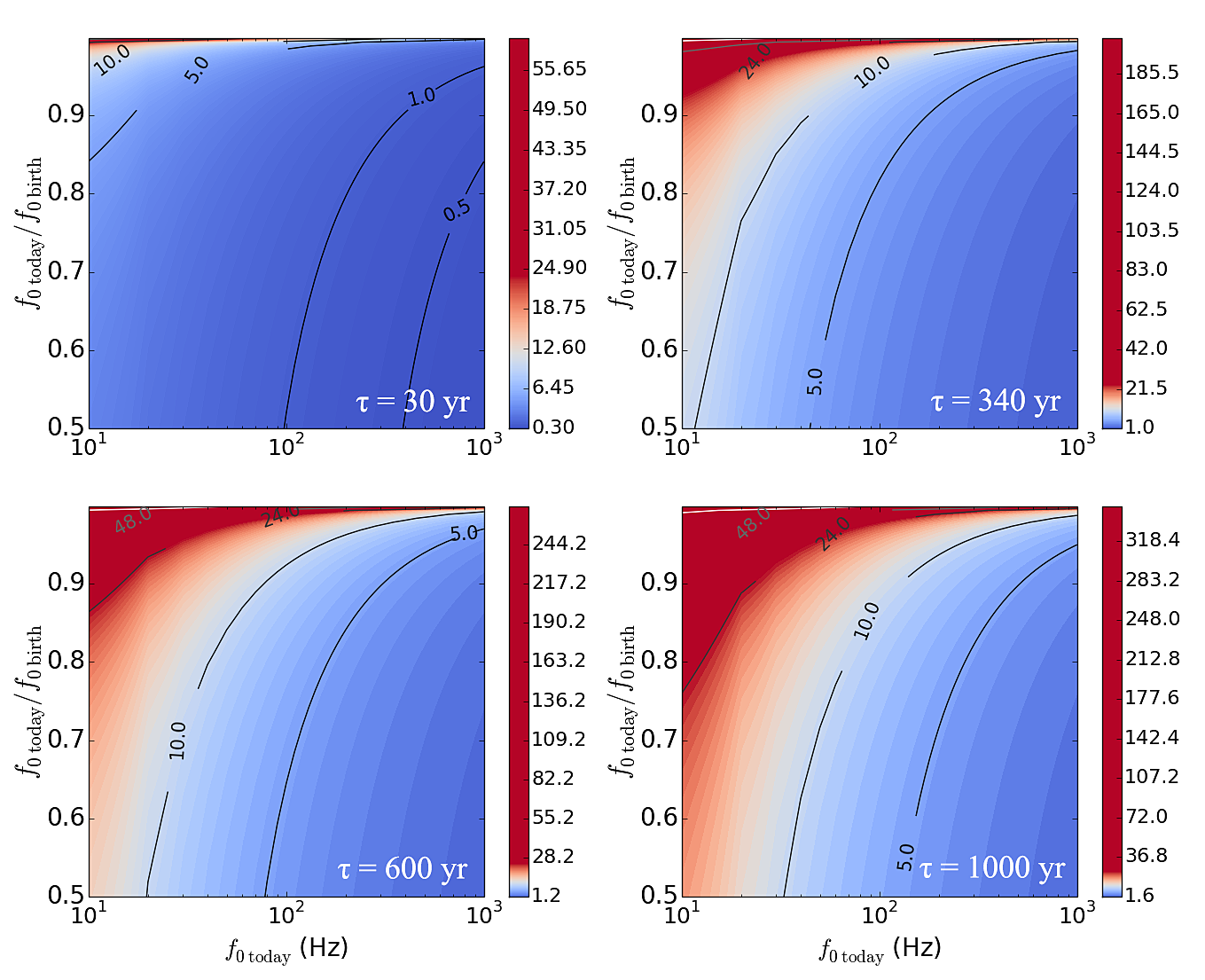}}
	\caption{Contours of $T_{\rm drift}$ (in hours) for targets with $\tau=0.03$\,kyr (top left), 0.34\,kyr (top right), 0.6\,kyr (bottom left) and 1\,kyr (bottom right) as a function of the gravitational-wave signal frequency today ${f_0}_{\rm today}$ and the ratio ${f_0}_{\rm today}/{f_0}_{\rm birth}$, where ${f_0}_{\rm birth}$ is the signal frequency at the birth of the star, assuming pure electromagnetic spin down ($\dot{f_0} \propto B_0^2f_0^n, n = 3$).}
	\label{fig:T_drift}
\end{figure*}

Given $\dot{f_0}$, we choose $T_{\rm drift}$ according to
\begin{equation}
\label{eqn:T_drift_form}
|\dot{f_0}| T_{\rm drift}=\Delta f_0=\frac{1}{2T_{\rm drift}}, 
\end{equation}
to satisfy (\ref{eqn:int_T_drift}). Hence $T_{\rm drift}=(2|\dot{f_0}|)^{-1/2}$ depends solely on the spin-down rate of the source. To illustrate how $T_{\rm drift}$ is determined by $\tau$, we calculate $\dot{f_0}$ and hence $T_{\rm drift}$ by assuming purely electromagnetic spin down ($\dot{f_0}(t) \propto B_0^2f_0(t)^n$, $n = 3$) as an example, where $B_0$ is the birth magnetic field strength; see Appendix~\ref{sec:estimate_T_drift} for a detailed derivation. Figure~\ref{fig:T_drift} displays contours of $T_{\rm drift}$ as a function of the gravitational-wave signal frequency today, ${f_0}_{\rm today}$, and the ratio ${f_0}_{\rm today}/{f_0}_{\rm birth}$, where ${f_0}_{\rm birth}$ is the signal frequency at the birth of the star. The contours in the upper left panel (e.g., SNR 1987A; $\tau = 0.03$\,kyr) show $T_{\rm drift}\lesssim 5$\,hr for most of the plot. By contrast, the contours in the upper right panel (e.g., Cas A; $\tau = 0.34$\,kyr) satisfy $T_{\rm drift} > 10$\,hr over $\approx 1/3$ area of the plot. The $T_{\rm drift}$ values estimated for older objects with $\tau=0.6$\,kyr and 1\,kyr are plotted in the lower panels. All panels show that $T_{\rm drift}$ decreases significantly for ${f_0}_{\rm today}/{f_0}_{\rm birth} \lesssim 0.8$ and ${f_0}_{\rm today}\gtrsim 100$\,Hz.

Figure~\ref{fig:T_drift} is plotted for $n=3$. In practice, the range of $\dot{f_0}$ is given by equation (\ref{eqn:dotf_range}) for $2\leq n \leq 7$. We substitute $|\dot{f_0}|_{\rm max}$ into (\ref{eqn:T_drift_form}) to satisfy (\ref{eqn:int_T_drift}) for all $\dot{f_0}$ and obtain
\begin{equation}
\label{eqn:T_drift_form_max}
T_{\rm drift} = (2|\dot{f_0}|_{\rm max})^{-1/2}.
\end{equation}
Hence for any given source with spin-down age $\tau$, we can always choose a coherent duration $T_{\rm drift}$ and divide the data into $N_T = T_{\rm obs}/T_{\rm drift}$ coherent segments, without searching the $(\dot{f_0},\ddot{f_0},\cdots)$ parameter space.

\subsection{Timing noise}
\label{sec:timning noise}
Stochastic spin wandering (often termed `timing noise') is a widespread phenomenon in isolated neutron stars pulsating at radio and X-ray wavelengths \cite{Hobbs2010,Shannon2010,Ashton2015}. The auto-correlation time-scale ranges from days to years \cite{Cordes1980,Price2012}. The phenomenon can result from magnetospheric changes \cite{Lyne2010}, superfluid dynamics in the stellar interior \cite{Alpar1986,Jones1990,Price2012,Melatos2014}, spin microjumps \cite{Cordes1985,Janssen2006}, and fluctuations in the spin-down torque \cite{Cheng1987,Cheng1987a,Urama2006}. 

In the absence of a measured ephemeris, a HMM can track the evolution of $f_0(t)$ caused by both secular spin down and stochastic timing noise \cite{Suvorova2016}. We approximate the timing noise by an unbiased random walk, in which $f_0(t)$ moves by at most one bin up or down during the timing-noise time-scale $T_{\rm drift}'$. The transition probability matrix for timing noise only is  
\begin{equation}
\label{eqn:tran_prob_sw}
A'_{q_{i-1} q_i} = A'_{q_i q_i} = A'_{q_{i+1} q_i}   = \frac{1}{3}.
\end{equation}
For $T_{\rm drift}' \gg T_{\rm drift}$, we can neglect timing noise, as discussed in previous sections. For $T_{\rm drift}' \ll T_{\rm drift}$, we can neglect secular spin down, set $A_{q_jq_i} = A'_{q_jq_i}$, and divide the data into $N_T = T_{\rm obs}/T_{\rm drift}'$ coherent segments. For $T_{\rm drift}' \approx T_{\rm drift}$, we choose $\min(T_{\rm drift}',T_{\rm drift})$ as the drift time-scale and adjust the transition probabilities to take into consideration both timing noise and spin down.

\section{Simulations and sensitivity}
\label{sec:simulations_and_sensi}
In this section, we firstly introduce the Viterbi score and the associated, score-based detection threshold in Section \ref{sec:viterbi_score}. We demonstrate the performance of $f_0$ tracking for three different scenarios using synthetic data: (1) an older object with weak timing noise (Section \ref{sec:1D-simulation-no-sw}); (2) an older object with strong timing noise (Section \ref{sec:1D-simulation-sw}); and (3) a very young object (Section \ref{sec:1D-young}). Simulations are conducted in an artificially restricted, 1-Hz sub-band to save time. The signals are injected at a fixed sky position, and $S_h(f)$ is set to Advanced LIGO's design sensitivity \cite{aLIGO_design_sensi}. We show the results in detail with a set of injections into Gaussian noise for each of the three scenarios, where the polarization and inclination angles and initial phase are arbitrarily chosen and fixed. Monte-Carlo simulations are conducted to generate the receiver operator characteristic (ROC) curves (Section \ref{sec:roc}).

\subsection{Viterbi score and threshold}
\label{sec:viterbi_score}
The Viterbi score $S$ is defined, such that the log likelihood of the optimal Viterbi path equals the mean log likelihood of all paths plus $S$ standard deviations at $t_{N_T}$, viz.
\begin{equation}
\label{eqn:viterbi_score}
S = \frac{\ln \delta_{q^*}{(t_{N_T})} -\mu_{\ln \delta}(t_{N_T})}{\sigma_{\ln \delta}(t_{N_T})}
\end{equation}
with
\begin{equation}
\mu_{\ln \delta}(t_{N_T}) = N_Q^{-1} \sum_{i=1}^{N_Q} \ln \delta_{q_i}(t_{N_T})
\end{equation}
and
\begin{equation}
\sigma_{\ln \delta}(t_{N_T})^2 = N_Q^{-1} \sum_{i=1}^{N_Q} [\ln \delta_{q_i}(t_{N_T}) - \mu_{\ln \delta}(t_{N_T}) ]^2,
\end{equation}
where $\delta_{q_i}(t_{N_T})$ denotes the maximum probability of the path ending in state $q_i$ ($1\leq i \leq N_Q$) at step $N_T$ (see Appendix \ref{sec:viterbi}), and $\delta_{q^*}{(t_{N_T})}$ is the likelihood of the optimal Viterbi path, i.e. $P[Q^*(O)|O]$. In a real search, we normally divide the full frequency band into multiple 1-Hz sub-bands to allow parallelized computing. In each 1-Hz sub-band, we consider the candidate for follow-up and further scrutiny, if $S$ exceeds a threshold $S_{\rm th}$ set by the desired false alarm and false dismissal probabilities. The value of $S_{\rm th}$ varies with $N_T$, $N_Q$, and the entries in $A_{q_j q_i}$. Systematic Monte-Carlo simulations are always required in practice to calculate $S_{\rm th}$ for each HMM implementation.

For the three scenarios in Sections \ref{sec:1D-simulation-no-sw}--\ref{sec:1D-young}, $S_{\rm th}$ is determined as follows. Searches are conducted on data sets containing pure Gaussian noise in 1-Hz sub-bands. For a given false alarm probability $P_{\rm a}$ in a 1-Hz sub-band, the value of $S$ yielding a fraction $P_{\rm a}$ of positive detections is $S_{\rm th}$. The false alarm probability in a search over band $B$ is given by $P_{\rm a, total} = 1- (1-P_{\rm a})^B$. We set $P_{\rm a} = 1\%$ and generate $10^3$ noise realizations for each scenario. Searches for the first two scenarios are based on the same $N_T$ and $N_Q$, and hence they both yield $S_{\rm th} = 6.7$. The mean and standard deviation of $S$ in the $10^3$ realizations are $\mu_S = 5.5$ and $\sigma_{S} = 0.4$. In the last scenario, we have $S_{\rm th} = 0.8$, with $\mu_S = 0.63$ and $\sigma_{S} = 0.06$. Because $\sigma_{\ln \delta}(t_{N_T})$ increases as $N_T$ gets larger, yielding lower $S$ normalized by $\sigma_{\ln \delta}(t_{N_T})$ in (\ref{eqn:viterbi_score}), it is as expected that the $S_{\rm th}$ is much lower in the last scenario ($N_T=2000$) than the first two scenarios ($N_T=40$).

\subsection{$\tau \gtrsim 5$\,kyr, $T_{\rm drift}' \gg T_{\rm drift}$}
\label{sec:1D-simulation-no-sw}

In the first group of tests, we consider a relatively older target with low timing noise, e.g., $\tau \gtrsim 5$\,kyr and $T_{\rm drift}' \gg T_{\rm drift}$. Four sets of synthetic data, containing injected signals with $h_0/10^{-26} =10$, 5, 3, and 2, are generated for $T_\text{obs}=83.3$\,d at two detectors (the LIGO Hanford and Livingston observatories) using \textit{Makefakedata} version 4 from LALApps. Detailed injection parameters are shown in Table \ref{tab:inj-1D-slow}. The searches are conducted using the search parameters in Table \ref{tab:search-1D-slow} and $A_{q_jq_i}$ in (\ref{eqn:trans_matrix}). The detection is deemed successful for $S>S_{\rm th}=6.7$. The results in Table \ref{tab:results-1D-slow} show that signals with $h_0 \geq 3 \times 10^{-26}$ are detected. We calculate the root-mean-square error (RMSE) $\varepsilon_{f_0}$ in $f_0$ between the optimal Viterbi path and the injected signal (in Hz and in units of $\Delta f_0$). All successful detections yield $\varepsilon_{f_0} < \Delta f_0$. The errors are introduced mostly because the HMM takes discrete values of $f_0$ (i.e., $\Delta f_0$ is the smallest step size), while the injected signal $f_0(t)$ can take any value within a bin.

\begin{figure*}
	\centering
	\subfigure[]
	{
		\label{fig:nosw-a}
		\scalebox{0.22}{\includegraphics{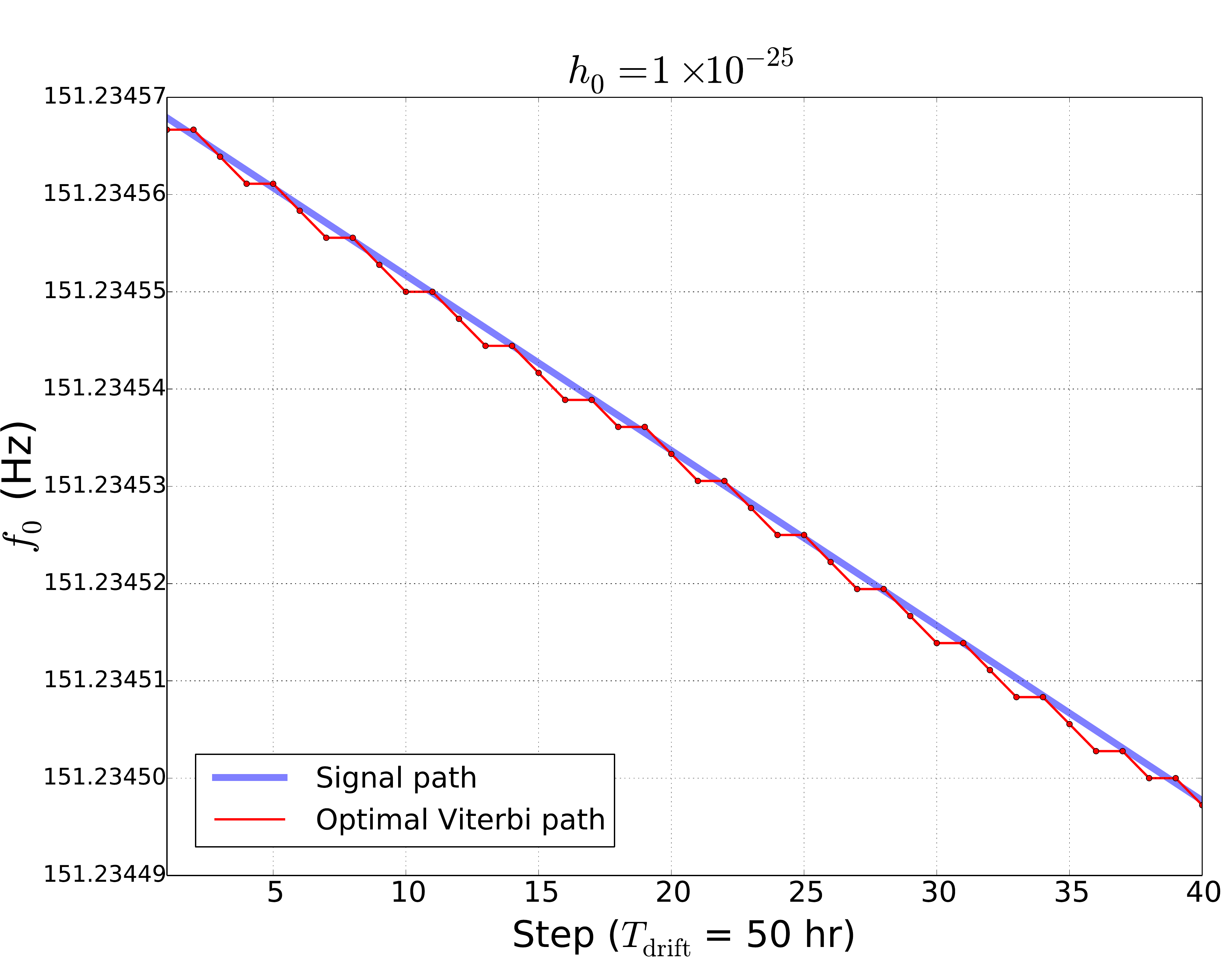}}
	}
	\subfigure[]
	{
		\label{fig:nosw-b}
		\scalebox{0.22}{\includegraphics{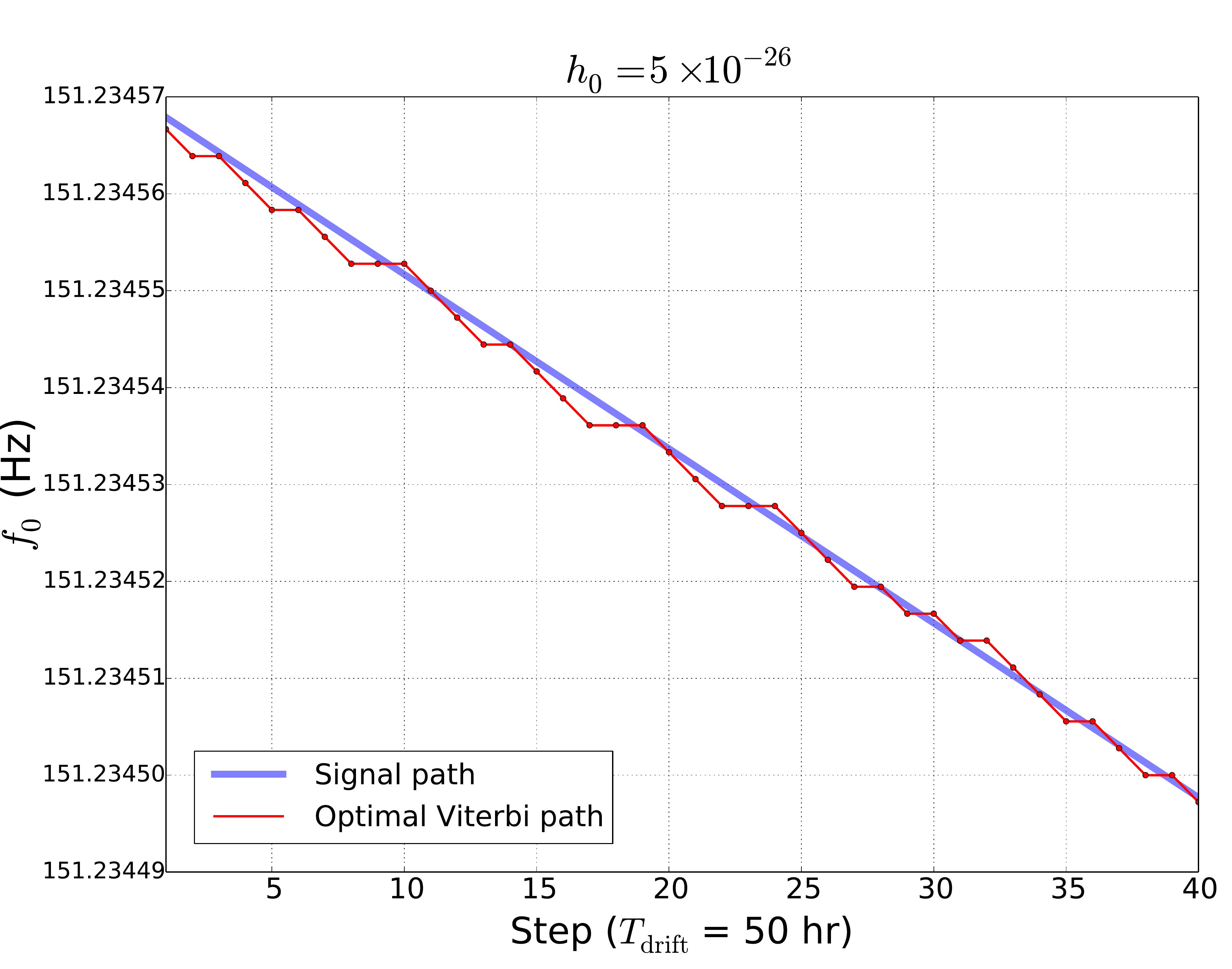}}
	}
	\subfigure[]
	{
		\label{fig:nosw-c}
		\scalebox{0.22}{\includegraphics{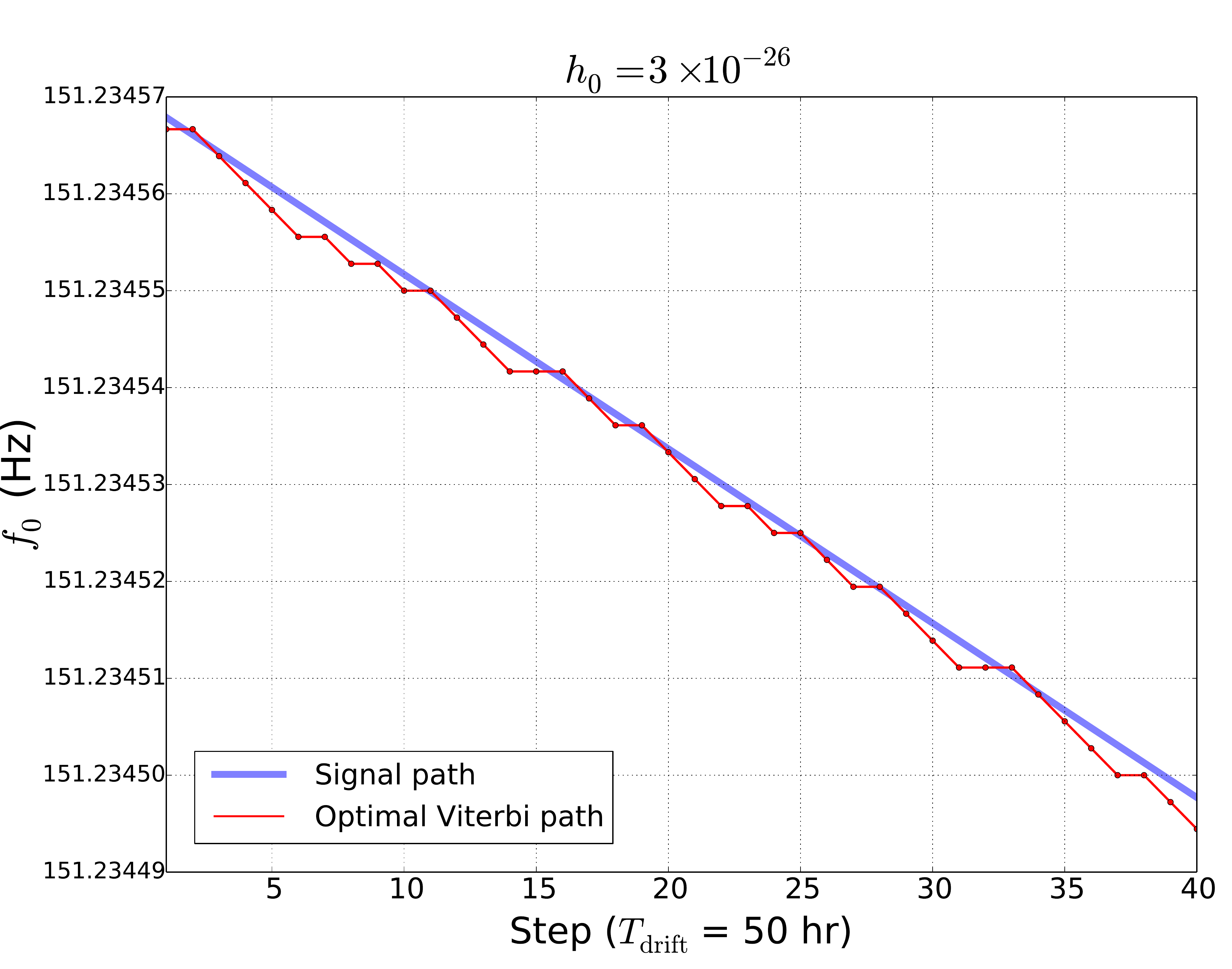}}
	}
	\subfigure[]
	{
		\label{fig:nosw-d}
		\scalebox{0.22}{\includegraphics{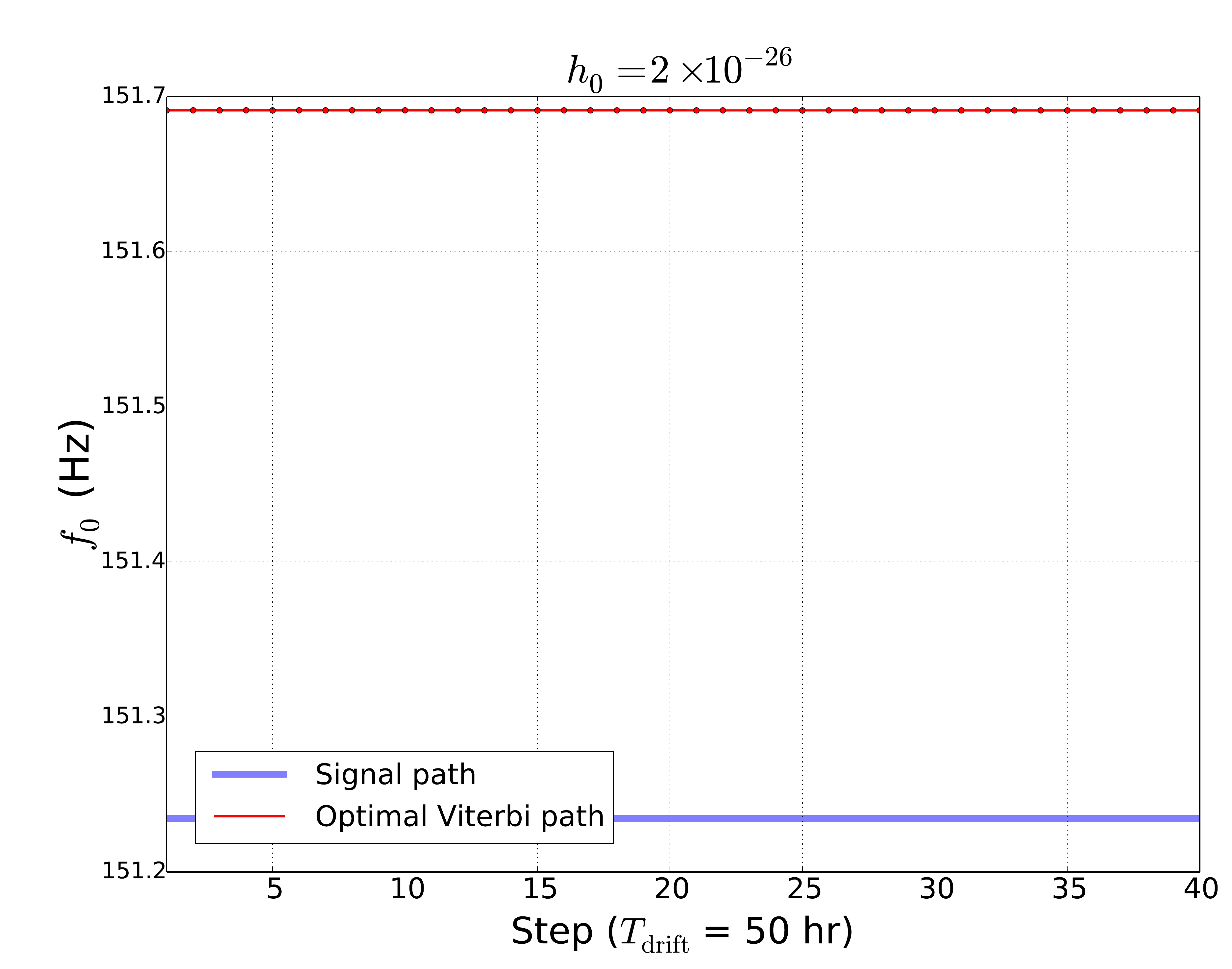}}
	}
	\caption{Injected $f_0(t)$ (blue curve) and optimal Viterbi path (red curve) for the injected signals in Table \ref{tab:results-1D-slow}. Panels (a)--(d) display paths for $h_0/10^{-26}=10$, 5, 3, 2, respectively. Good matches are obtained in (a)--(c), with $\varepsilon_{f_0} = 0.39 \Delta f_0$, $0.46 \Delta f_0$, and $0.59 \Delta f_0$, respectively. In (d), the signal is not detected; the spin down of $f_0(t)$ is too slow to be seen in the plot ($\varepsilon_{f_0} \gg \dot{f_0}T_{\rm obs}= 7\times 10^{-5}$\,Hz). The horizontal axes are in units of HMM steps with $T_{\rm drift} = 50$\,hr for each step ($N_T = 40$).}
	\label{fig:1D_2d_tracking_results}
\end{figure*}

Figure \ref{fig:1D_2d_tracking_results} presents the tracking results corresponding to Table \ref{tab:results-1D-slow}. Panels (a)--(c) show that the optimal Viterbi paths match the injected $f_0(t)$ closely, with $\varepsilon_{f_0} = 0.39 \Delta f_0$, $0.46 \Delta f_0$ and $0.59 \Delta f_0$, respectively. Panel (d) shows that the signal is not tracked successfully. The detectability drops rapidly from $h_0= 3 \times 10^{-26}$ to $h_0 = 2 \times 10^{-26}$, as expected near the detection limit (see detailed explanation in Section III B of Ref. \cite{Suvorova2016}).

\begin{table}
	\centering
	\setlength{\tabcolsep}{3pt}
	\renewcommand\arraystretch{1.4}
	\begin{tabular}{lll}
		\hline
		\hline
		Parameter & Symbol & Value \\
		\hline
		Right ascension & $\alpha$ & 23h 23m 26.0s\\
		Declination & $\delta$ & $58^{\circ}48' 0.0''$\\
		Polarization angle &$\psi$ & 4.94278\,rad\\
		Inclination angle &$\cos\iota$ & 0.718742 \\
		Initial phase &$\phi_0$ & 2.43037\,rad\\
		PSD &$S_h (f)^{1/2}$ & $4 \times 10^{-24}$\,Hz$^{-1/2}$ \\
		Frequency & ${f_0}_{\rm inj}$ & 151.23456789\,Hz \\
		First derivative of ${f_0}_{\rm inj}$ & ${\dot{f_0}}_{\rm inj}$ & $-1.0 \times 10^{-11}$\,Hz\,s$^{-1}$ \\
		Second derivative of ${f_0}_{\rm inj}$ & ${\ddot{f_0}}_{\rm inj}$ & $2.0 \times 10^{-24}$\,Hz\,s$^{-2}$ \\
		\hline
		\hline
	\end{tabular}
	\caption[parameters]{Injection parameters used to create the synthetic data analyzed in Sections \ref{sec:1D-simulation-no-sw} and \ref{sec:1D-simulation-sw}.}
	\label{tab:inj-1D-slow}
\end{table}

\begin{table}
	\centering
	\setlength{\tabcolsep}{4pt}
	\renewcommand\arraystretch{1.4}
	\begin{tabular}{lll}
		\hline
		\hline
		Parameter & Value & Unit\\
		\hline
		$f_0$ &151--152 & Hz \\
		$T_{\rm drift}$ & 50 & hr \\
		$\Delta f_0$ & $2.78 \times 10^{-6}$ & Hz \\
		$T_{\rm obs}$ & 83.3 & d\\
		$N_T$ & 40 & --\\
		\hline
		\hline
	\end{tabular}
	\caption[parameters]{Search parameters for the synthetic signals with injection parameters quoted in Table \ref{tab:inj-1D-slow}.}
	\label{tab:search-1D-slow}
\end{table}

\begin{table}
	\centering
	\setlength{\tabcolsep}{5pt}
	\renewcommand\arraystretch{1.4}
	\begin{tabular}{lllll}
		\hline
		\hline
		$h_0$ $(10^{-26})$ & Detect? & $S$& $\varepsilon_{f_0}$ (Hz) & $\varepsilon_{f_0}/\Delta f_0$\\
		\hline
		$10.0$ & $\checkmark$ & 90.9 & $1.07\times 10^{-6}$ & 0.39\\
		$5.0$ & $\checkmark$  & 32.0 &$1.29\times 10^{-6}$ & 0.46  \\
		$3.0$ & $\checkmark$  & 10.6 &$1.64\times 10^{-6}$ & 0.59  \\
		$2.0$ & $\times$  & 5.5 & 0.46 & $1.6\times 10^5$ \\
		\hline
		\hline
	\end{tabular}
	\caption[]{Results of $f_0$ tracking for synthetic signals with the injection parameters in Table \ref{tab:inj-1D-slow}, $T_\text{obs}=83.3$\,d, $T_\text{drift}=50$\,hr, and characteristic wave strain $h_0$. The RMSE $\varepsilon_{f_0}$ between the optimal Viterbi path and the injected $f_0(t)$ is quoted in Hz and in units of $\Delta f_0 = 2.78 \times 10^{-6}$\,Hz. The third column quotes the Viterbi score $S$.}
	\label{tab:results-1D-slow}
\end{table}

\subsection{$\tau \gtrsim 5$\,kyr, $T_{\rm drift}' \approx T_{\rm drift}$}
\label{sec:1D-simulation-sw}

In the second group of tests, we show that the HMM can track secular spin down and timing noise simultaneously for synthetic signals injected in Gaussian noise. As an example, we assume the time-scale of the unbiased random walk is the same as the spin-down time-scale, i.e., $T_{\rm drift}' = T_{\rm drift} = 50$\,hr. The modified transition probability matrix is the product of (\ref{eqn:trans_matrix}) and (\ref{eqn:tran_prob_sw}), given by
\begin{equation}
\label{eqn:trans_matrix_3_1}
2A_{q_{i-2} q_i} = A_{q_{i-1} q_i}=A_{q_i q_i}=2A_{q_{i+1} q_i}= \frac{1}{3}, 
\end{equation}
with all other terms being zero. 

Four data sets with signal strains $h_0/10^{-26} =10$, $5$, $3$ and $2$ are generated for $T_\text{obs}=83.3$\,d at two detectors. We use the same injection parameters in Table \ref{tab:inj-1D-slow} at $t=0$. In addition to the spin down, $f_0(t)$ wanders randomly by at most $\pm \Delta f_0$ over time-scale $T_{\rm drift}$. The data sets are searched using the parameters in Table \ref{tab:search-1D-slow} and $A_{q_jq_i}$ in (\ref{eqn:trans_matrix_3_1}). The results are shown in Table \ref{tab:results-1D-slow-sw}. The detection is deemed successful for $S>S_{\rm th}=6.7$. All successful detections yield $\varepsilon_{f_0} < \Delta f_0$.

\begin{figure*}
	\centering
	\subfigure[]
	{	
		\label{fig:sw-a}
		\scalebox{0.22}{\includegraphics{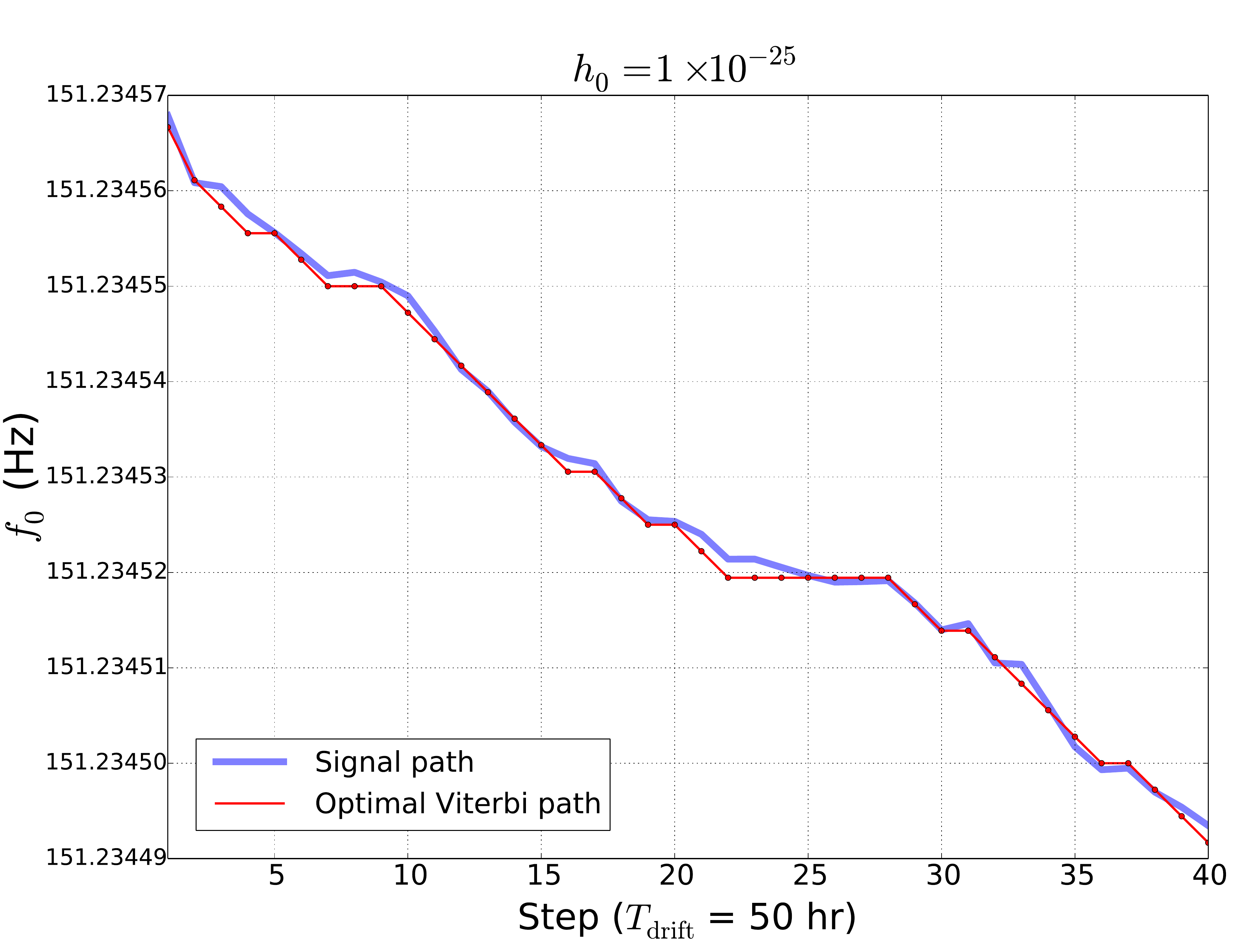}}
	}
	\subfigure[]
	{
		\label{fig:sw-b}
		\scalebox{0.22}{\includegraphics{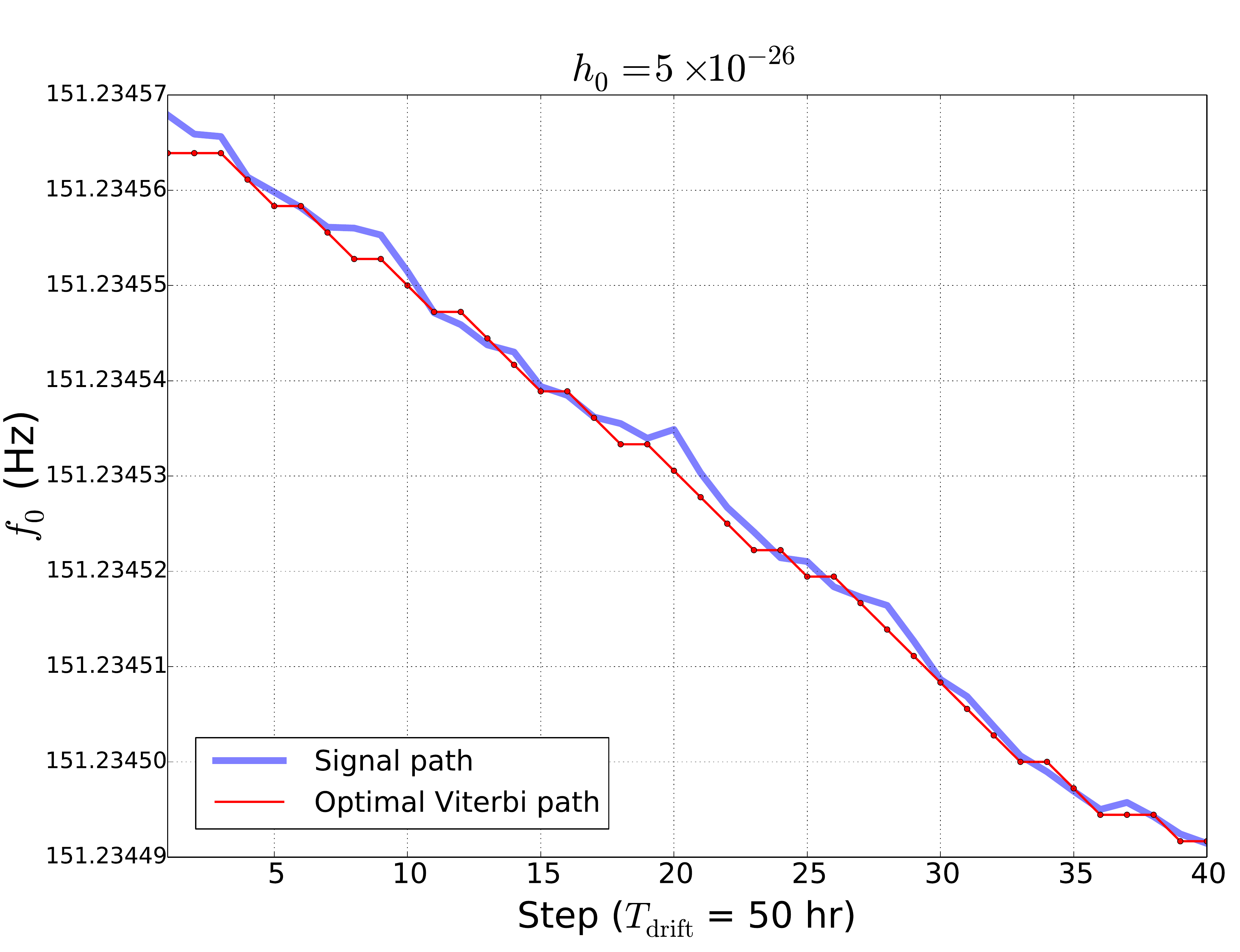}}
	}
	\subfigure[]
	{
		\label{fig:sw-c}
		\scalebox{0.22}{\includegraphics{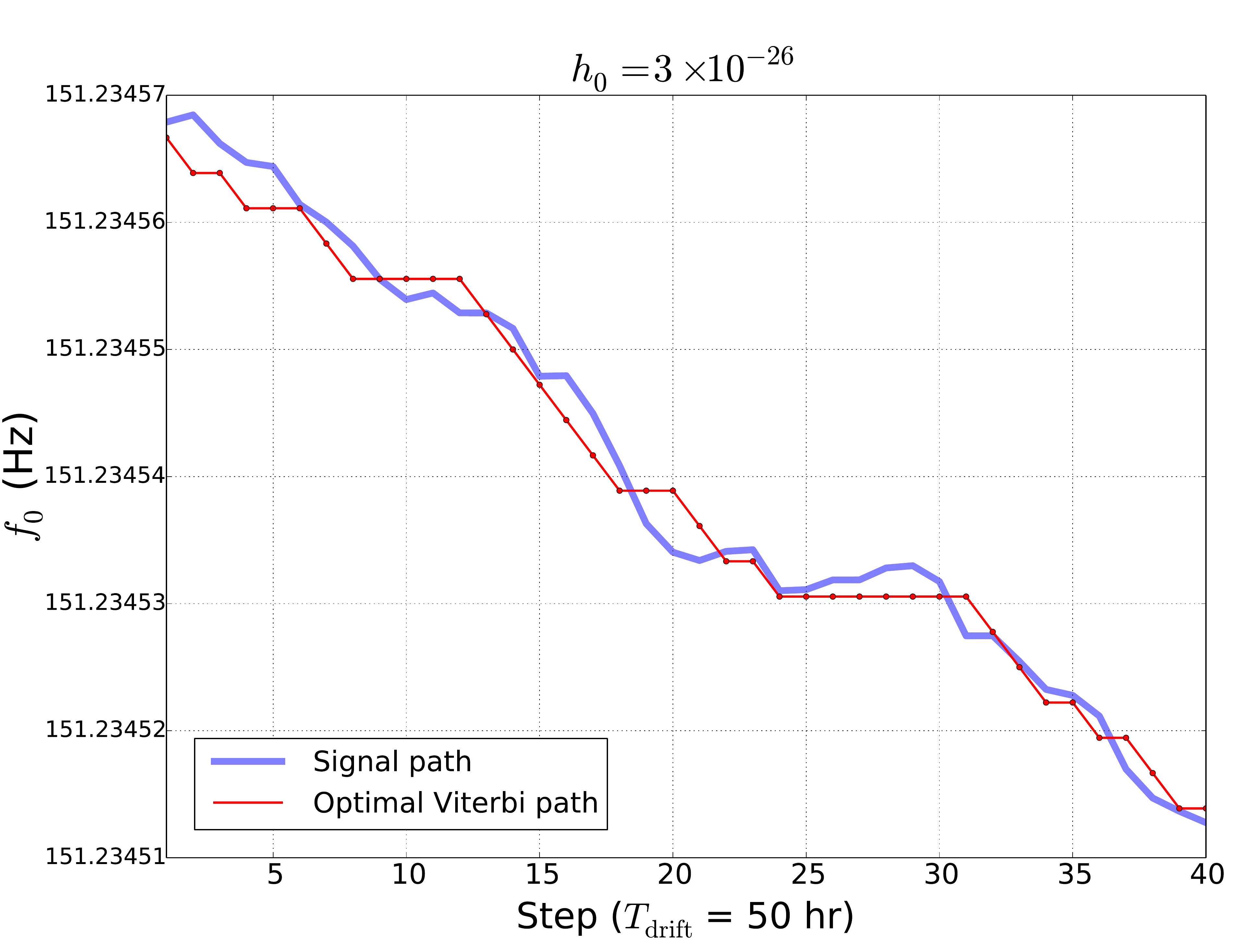}}
	}
	\subfigure[]
	{
		\label{fig:sw-d}
		\scalebox{0.22}{\includegraphics{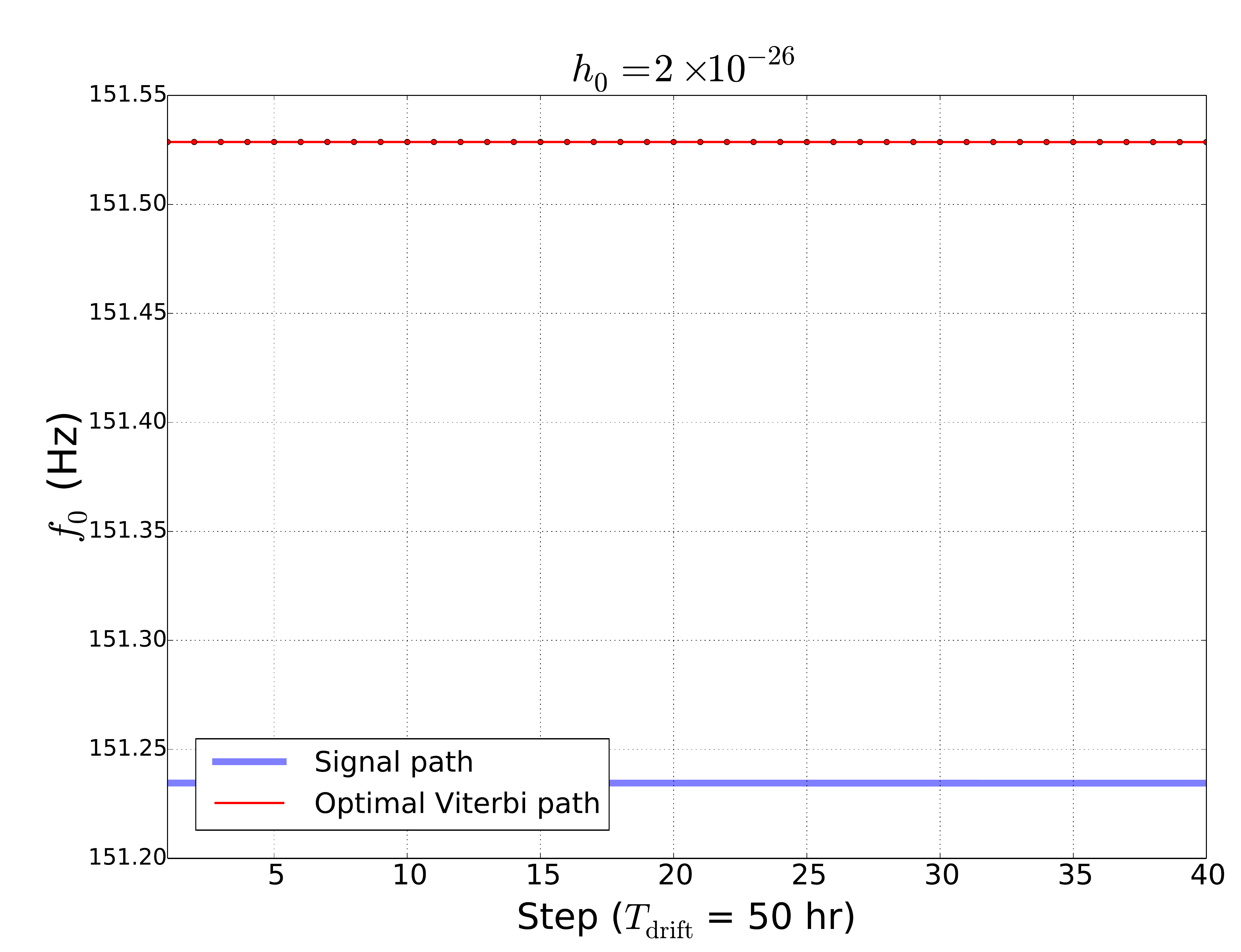}}
	}
	\caption{Injected $f_0(t)$ (blue curve) and optimal Viterbi path (red curve) for the injected signals in Table \ref{tab:results-1D-slow-sw}. Panels (a)--(d) display paths for $h_0/10^{-26}=10$, 5, 3, 2, respectively. Good matches are obtained in (a)--(c), with $\varepsilon_{f_0} = 0.38 \Delta f_0$, $0.59 \Delta f_0$, and $0.77 \Delta f_0$, respectively. In (d), the signal is not detected; the spin down of $f_0(t)$ is too slow to be seen in the plot ($\varepsilon_{f_0} \gg \dot{f_0}T_{\rm obs}\sim 7\times 10^{-5}$\,Hz). The horizontal axes are in units of HMM steps with $T_{\rm drift} = 50$\,hr for each step ($N_T = 40$).}
	\label{fig:1D_sw_tracking_results}
\end{figure*}

Figure \ref{fig:1D_sw_tracking_results} presents the tracking results for signals with $h_0/10^{-26} = 10$, 5, 3, 2 in Table \ref{tab:results-1D-slow-sw}. The optimal Viterbi paths in panels (a)--(c) match the injected paths closely, indicating successful detections. The RMSE $\varepsilon_{f_0}$ increases from $0.38 \Delta f_0$ to $0.77 \Delta f_0$, when $h_0$ decreases from $1\times 10^{-25}$ to $3\times 10^{-26}$. Panel (d) shows that the optimal Viterbi path does not match the injected $f_0(t)$ for $h_0 = 2\times 10^{-26}$, i.e., the injected signal is not detected. 

\begin{table}
	\centering
	\setlength{\tabcolsep}{3pt}
	\renewcommand\arraystretch{1.4}
	\begin{tabular}{lllll}
		\hline
		\hline
		$h_0$ $(10^{-26})$ & Detect? & $S$& $\varepsilon_{f_0}$ (Hz) & $\varepsilon_{f_0}/\Delta f_0$\\
		\hline
		$10.0$ & $\checkmark$ & 92.5 & $1.06\times 10^{-6}$ & 0.38\\
		$5.0$ & $\checkmark$  & 28.4 &$1.63\times 10^{-6}$ & 0.59  \\
		$3.0$ & $\checkmark$  & 7.2 &$2.14\times 10^{-6}$ & 0.77  \\
		$2.0$ & $\times$  & 5.1 & 0.29 & $1.1\times 10^5$ \\
		\hline
		\hline
	\end{tabular}
	\caption[]{Results of $f_0$ tracking for synthetic spin-wandering signals with the injection parameters in Table \ref{tab:inj-1D-slow}, $T_\text{obs}=83.3$\,d, $T_\text{drift}=50$\,hr, and characteristic wave strain $h_0$. The RMSE $\varepsilon_{f_0}$ between the optimal Viterbi path and the injected $f_0(t)$ is quoted in Hz and in units of $\Delta f_0 = 2.78 \times 10^{-6}$\,Hz. The third column quotes the Viterbi score $S$.}
	\label{tab:results-1D-slow-sw}
\end{table}

Given the signal-to-noise ratio, $T_{\rm drift}$ and $N_T$, the sensitivity of the search remains the same for $T_{\rm drift}' \gg T_{\rm drift}$ (Section \ref{sec:1D-simulation-no-sw}) and $T_{\rm drift}' = T_{\rm drift}$ (Section \ref{sec:1D-simulation-sw}). 

\subsection{$\tau \lesssim 0.03$\,kyr, $T_{\rm drift}' \gg T_{\rm drift}$}
\label{sec:1D-young}

In the third group of tests, we consider a very young object with $\tau \lesssim 0.03$\,kyr, e.g., SNR 1987A. In stack-slide-based searches for such young sources, typically four or more frequency derivatives must be searched in order to accurately track the rapid phase evolution. 

\begin{table}
	\centering
	\setlength{\tabcolsep}{3pt}
	\renewcommand\arraystretch{1.4}
	\begin{tabular}{lll}
		\hline
		\hline
		Parameter & Symbol & Value \\
		\hline
		First derivative of ${f_0}_{\rm inj}$ & ${\dot{f_0}}_{\rm inj}$ & $-3.0 \times 10^{-8}$\,Hz\,s$^{-1}$ \\
		Second derivative of ${f_0}_{\rm inj}$ & ${\ddot{f_0}}_{\rm inj}$ & $3.0 \times 10^{-17}$\,Hz\,s$^{-2}$ \\
		Third derivative of ${f_0}_{\rm inj}$ & ${\dddot{f_0}}_{\rm inj}$ & $-3.0 \times 10^{-26}$\,Hz\,s$^{-3}$ \\
		\hline
		\hline
	\end{tabular}
	\caption[parameters]{Spin-down-related injection parameters used to create the synthetic data analyzed in Section \ref{sec:1D-young}.}
	\label{tab:inj-fast-sd}
\end{table}

We inject four signals with $h_0/10^{-26}=15$, 13, 11, 10 and high spin-down rates as quoted in Table \ref{tab:inj-fast-sd}. Other injection parameters remain the same as those in Table \ref{tab:inj-1D-slow}. We choose $T_{\rm drift} = 1$\,hr to satisfy equation (\ref{eqn:int_T_drift}). In this case, we always have $T_{\rm drift}' \gg T_{\rm drift}$ and hence use $A_{q_jq_i}$ in (\ref{eqn:trans_matrix}). The search parameters and results are presented in Tables \ref{tab:search-1D-fast} and \ref{tab:results-1D-fast}, respectively. In this group, the detection is deemed successful for $S > S_{\rm th} =0.8$ given $P_{\rm a}=1\%$. The successful detections yield $\varepsilon_{f_0} < 2\Delta f_0 = 2.8 \times 10^{-4}$\,Hz. We tolerate $\varepsilon_{f_0}$ slightly larger than $\Delta f_0$ because $T_{\rm drift}$ is relatively short. 

\begin{figure*}
	\centering
	\subfigure[]
	{
		\label{fig:1d-2k-a}
		\scalebox{0.26}{\includegraphics{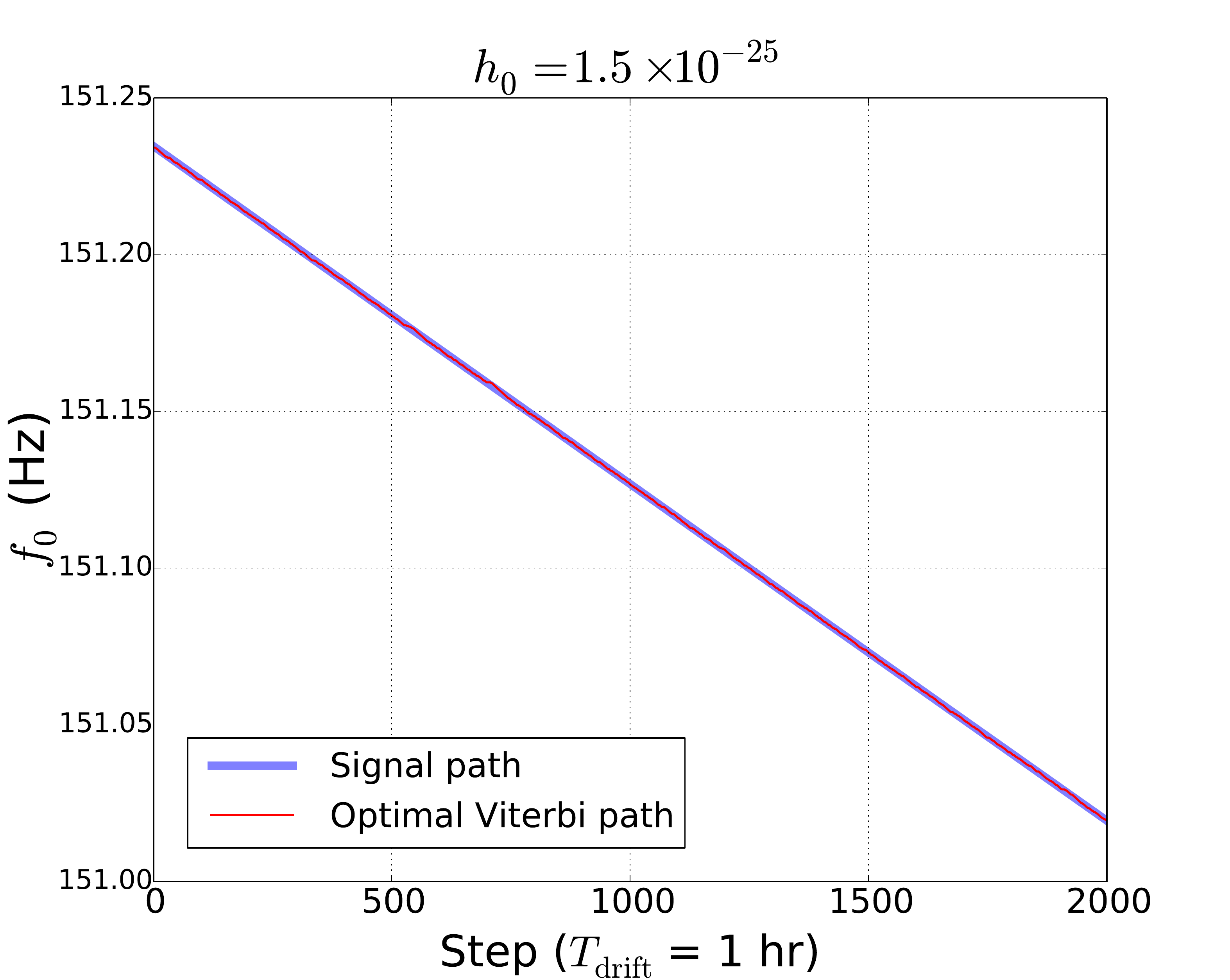}}
	}
	\subfigure[]
	{
		\label{fig:1d-2k-b}
		\scalebox{0.26}{\includegraphics{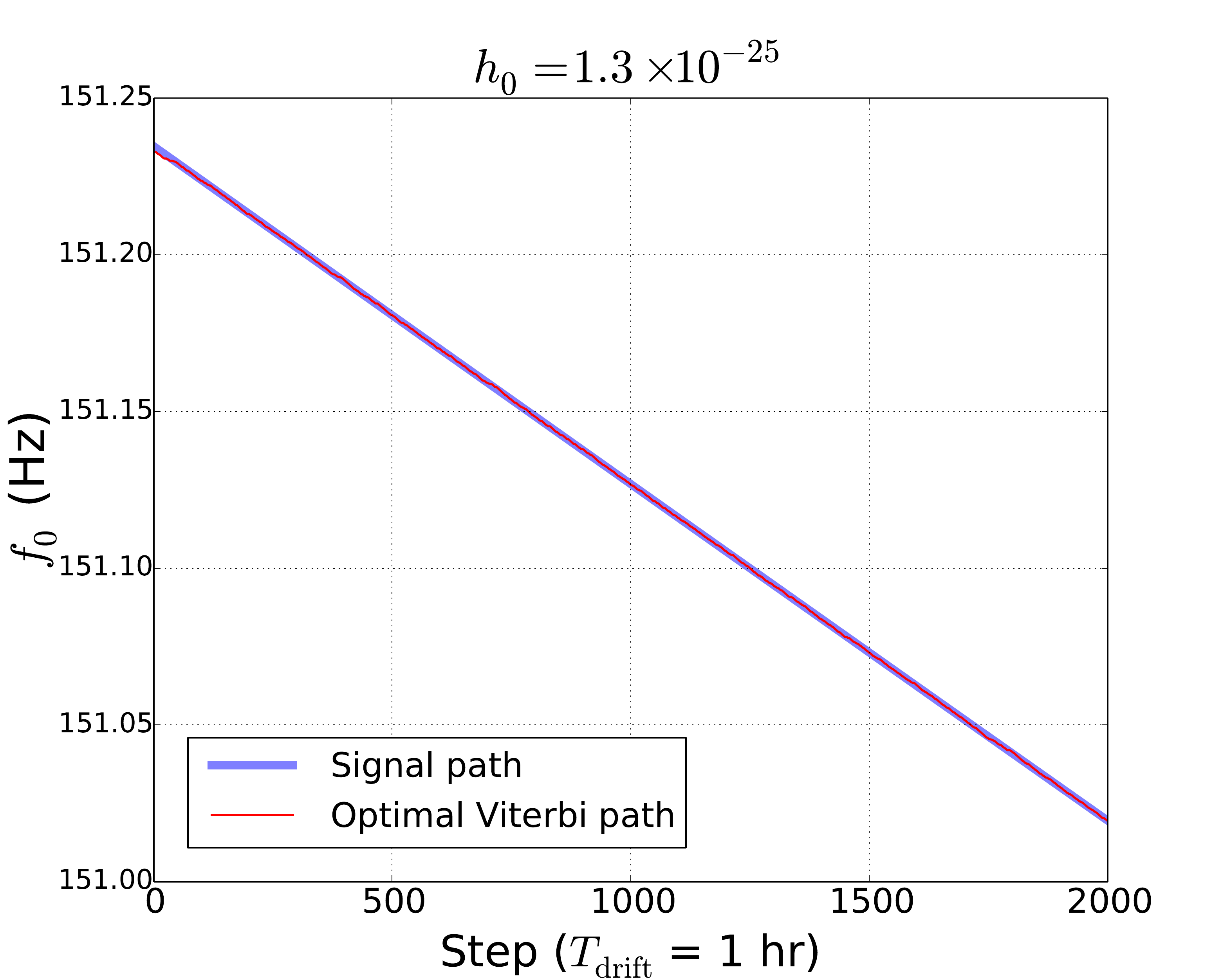}}
	}
	\subfigure[]
	{
		\label{fig:1d-2k-c}
		\scalebox{0.26}{\includegraphics{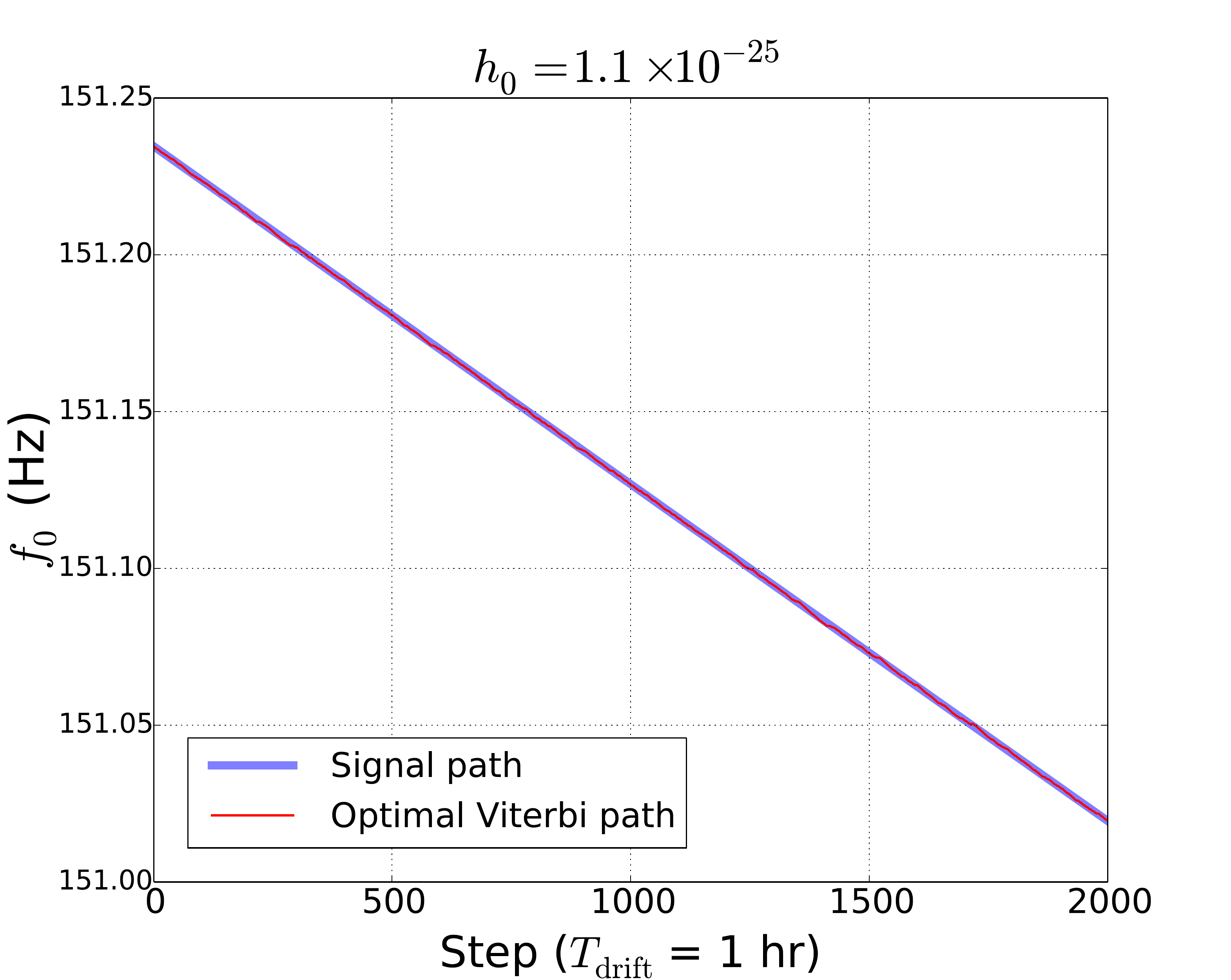}}
	}
	\subfigure[]
	{
		\label{fig:1d-2k-d}
		\scalebox{0.26}{\includegraphics{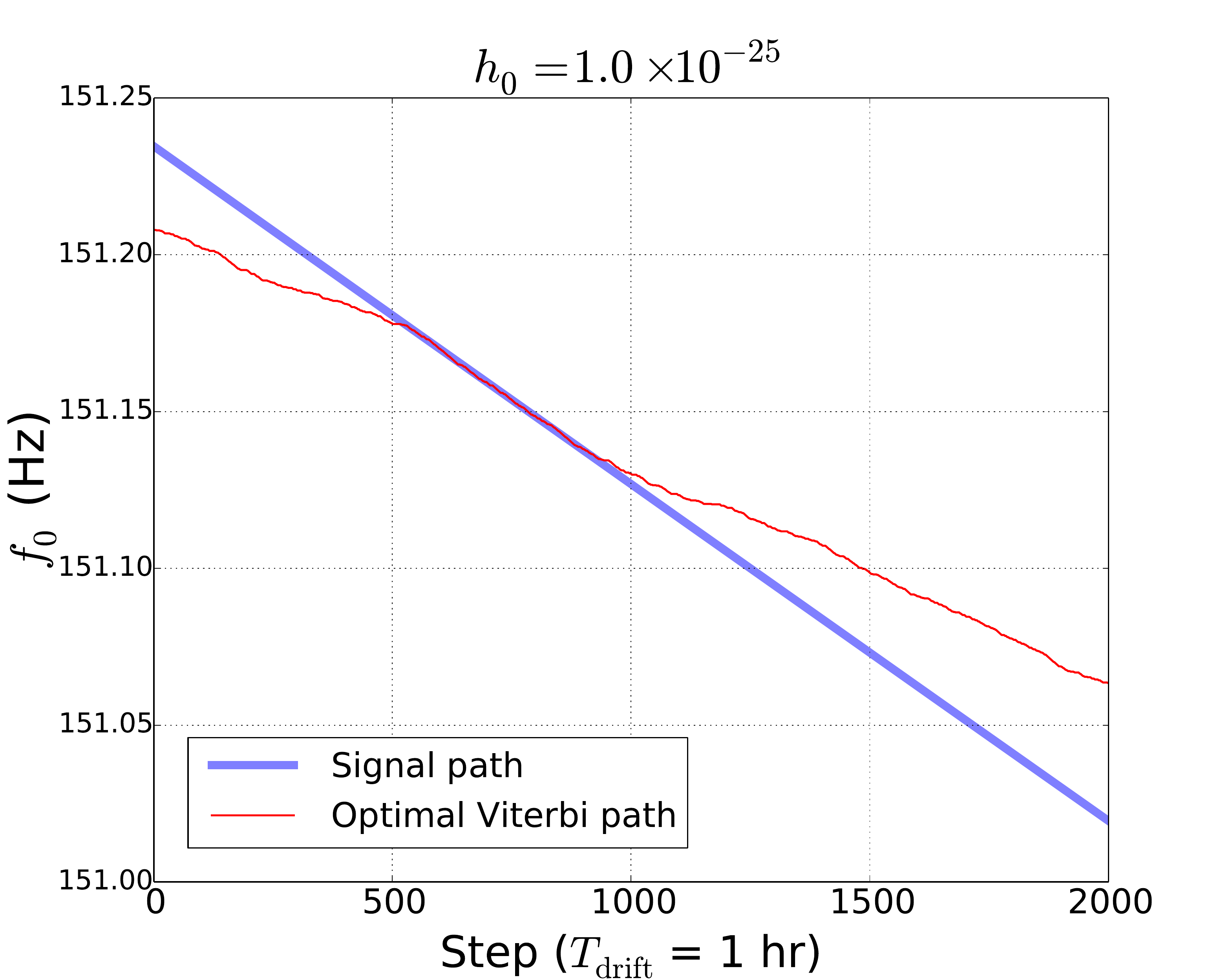}}
	}
	\caption{Injected $f_0(t)$ (blue curve) and optimal Viterbi path (red curve) for the injected signals in Table \ref{tab:results-1D-fast}. Panels (a)--(d) display paths for $h_0/10^{-26}=15$, 13, 11, 10, respectively. Good matches are obtained in (a)--(c), with $\varepsilon_{f_0} <2 \Delta f_0$. In panel (d), the signal is not detected. The horizontal axes are in units of HMM steps with $T_{\rm drift} = 1$\,hr for each step ($N_T = 2000$).}
	\label{fig:1D_2k_tracking_results}
\end{figure*}

Figure \ref{fig:1D_2k_tracking_results} shows the tracking results corresponding to Table \ref{tab:results-1D-fast}. Panels (a)--(c) show that the optimal Viterbi paths match the injected $f_0(t)$ closely. The discrepancy between the optimal Viterbi path and the injected $f_0(t)$ can hardly be seen, because $\varepsilon_{f_0}  \sim 10^{-4}$\,Hz is much smaller than the total change in $f_0$ over $T_{\rm obs}$ ($\approx 0.2$\,Hz). Panel (d) shows that the signal is not detected for $h_0 = 10 \times 10^{-26}$, with $\varepsilon_{f_0}=151.8 \Delta f_0 \gg \Delta f_0$.

\begin{table}
	\centering
	\setlength{\tabcolsep}{4pt}
	\renewcommand\arraystretch{1.4}
	\begin{tabular}{lll}
		\hline
		\hline
		Parameter & Value & Unit\\
		\hline
		$f_0$ &151--152 & Hz \\
		$T_{\rm drift}$ & 1 & hr \\
		$\Delta f_0$ & $1.39 \times 10^{-4}$ & Hz \\
		$T_{\rm obs}$ & 83.3 & d\\
		$N_T$ & 2000 & --\\
		\hline
		\hline
	\end{tabular}
	\caption[parameters]{Search parameters for the synthetic signals with injection parameters quoted in Tables \ref{tab:inj-1D-slow} and \ref{tab:inj-fast-sd}.}
	\label{tab:search-1D-fast}
\end{table}

\begin{table}
	\centering
	\setlength{\tabcolsep}{5pt}
	\renewcommand\arraystretch{1.4}
	\begin{tabular}{lllll}
		\hline
		\hline
		$h_0$ $(10^{-26})$ & Detect? & $S$& $\varepsilon_{f_0}$ (Hz) & $\varepsilon_{f_0}/\Delta f_0$\\
		\hline
		$15.0$ & $\checkmark$ & 3.0 & $1.77\times 10^{-4}$ & 1.3 \\
		$13.0$ & $\checkmark$  & 2.1 &$2.50\times 10^{-4}$ &  1.8 \\
		$11.0$ & $\checkmark$  & 0.9 & $2.47\times 10^{-4}$ &  1.8\\
		$10.0$ & $\times$  & 0.5 & 0.02 & 151.8\\
		\hline
		\hline
	\end{tabular}
	\caption[]{Results of $f_0$ tracking for synthetic signals with with the injection parameters in Tables \ref{tab:inj-1D-slow} and \ref{tab:inj-fast-sd}, $T_\text{obs}=83.3$\,d, $T_\text{drift}=1$\,hr, and characteristic wave strain $h_0$. The RMSE $\varepsilon_{f_0}$ between the optimal Viterbi path and the injected $f_0(t)$ is quoted in Hz and in units of $\Delta f_0 = 1.39 \times 10^{-4}$\,Hz. The third column quotes the Viterbi score $S$.}
	\label{tab:results-1D-fast}
\end{table}

\subsection{ROC curve and sensitivity}
\label{sec:roc}
The detection threshold $S_{\rm th}$ is set by $P_{\rm a}$. The probability that an injected signal yields $S < S_{\rm th}$ is the false dismissal probability, denoted by $P_{\rm d}$. We quantify the performance of the HMM in terms of its ROC curve, plotting the detection probability $1-P_{\rm d}$ against the false alarm probability $P_{\rm a}$ for various signal strengths. The signal-to-noise ratio for a biaxial rotor scales approximately in proportion to $h_0^{\rm eff}$, given by \cite{Jaranowski1998,Messenger2015}
\begin{equation}
\label{eqn:h0eff}
h_0^{\rm eff}= h_0\,2^{-1/2}\{{[(1+\cos^2\iota)/2]^2 + \cos^2\iota}\}^{1/2}, 
\end{equation}
so we quote $h_0^{\rm eff}$ instead of $h_0$ as the signal strength. The simulations are conducted in an artificially restricted, 1-Hz sub-band, at a fixed sky location, with both polarization angle $\psi$ and initial phase $\Phi_0$ randomly chosen with a uniform distribution within the range $[0,2\pi]$\,rad.

\begin{figure}
	\centering
	\scalebox{0.38}{\includegraphics{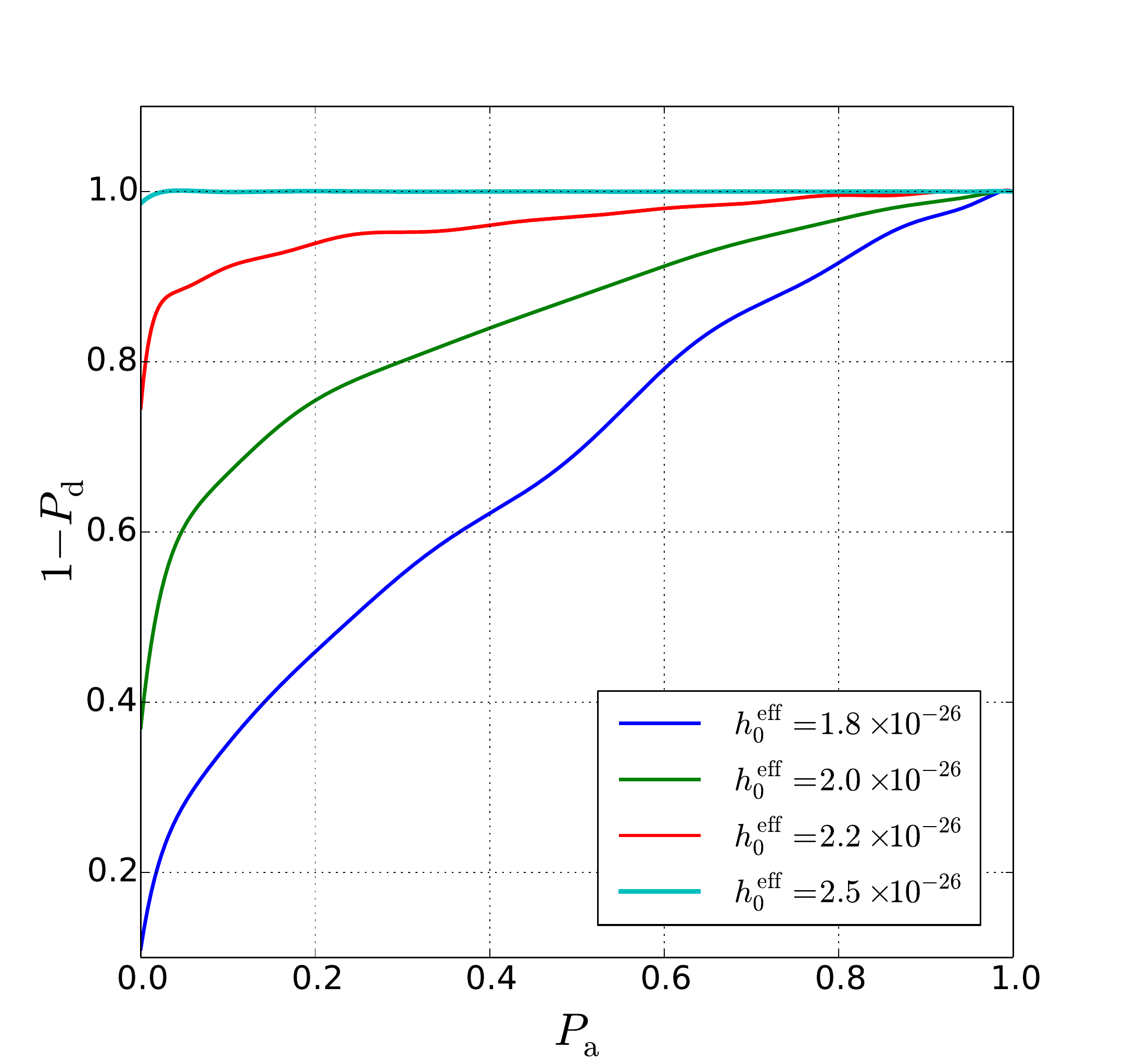}}
	\caption{Receiver operator characteristic (ROC) curves for the searches described in Section \ref{sec:1D-simulation-no-sw} and \ref{sec:1D-simulation-sw} (curves indistinguishable to the eye). The four curves (from top to bottom) correspond to the four representative wave strains $h_0^{\rm eff}/10^{-26} = 2.5$, 2.2, 2.0, and 1.8. The horizontal and vertical axes indicate the false alarm probability $P_{\rm a}$ and detection probability $1-P_{\rm d}$, respectively. Each curve is based on 500 realizations. Parameters: $T_{\rm drift} =50$\,hr, $N_T = 40$, $S_h (f)^{1/2} = 4 \times 10^{-24}$\,Hz$^{-1/2}$.}
	\label{fig:roc1}
\end{figure}

The ROC curves are essentially indistinguishable for the two scenarios in Section \ref{sec:1D-simulation-no-sw} and \ref{sec:1D-simulation-sw}, because HMM tracking is insensitive to the exact choice of $A_{q_jq_i}$ \cite{Quinn2001,Viterbi1967}. Figure \ref{fig:roc1} shows the ROC curves for these two scenarios with four values of $h_0^{\rm eff}$, ranging from $1.8 \times 10^{-26}$ to $2.5 \times 10^{-26}$. For $P_{\rm a}=1\%$, we have $85\%$ and $99\%$ confidence to detect a signal with $h_0^{\rm eff}= 2.2 \times 10^{-26}$ and $2.5 \times 10^{-26}$, respectively, read off the top two curves in Figure \ref{fig:roc1}. The $95\%$ confidence sensitivity on effective strain is $h_0^{\rm eff,95\%} \approx 2.4 \times 10^{-26}$ ($T_{\rm obs} = 83.3$\,d).

\begin{figure}
	\centering
	\scalebox{0.38}{\includegraphics{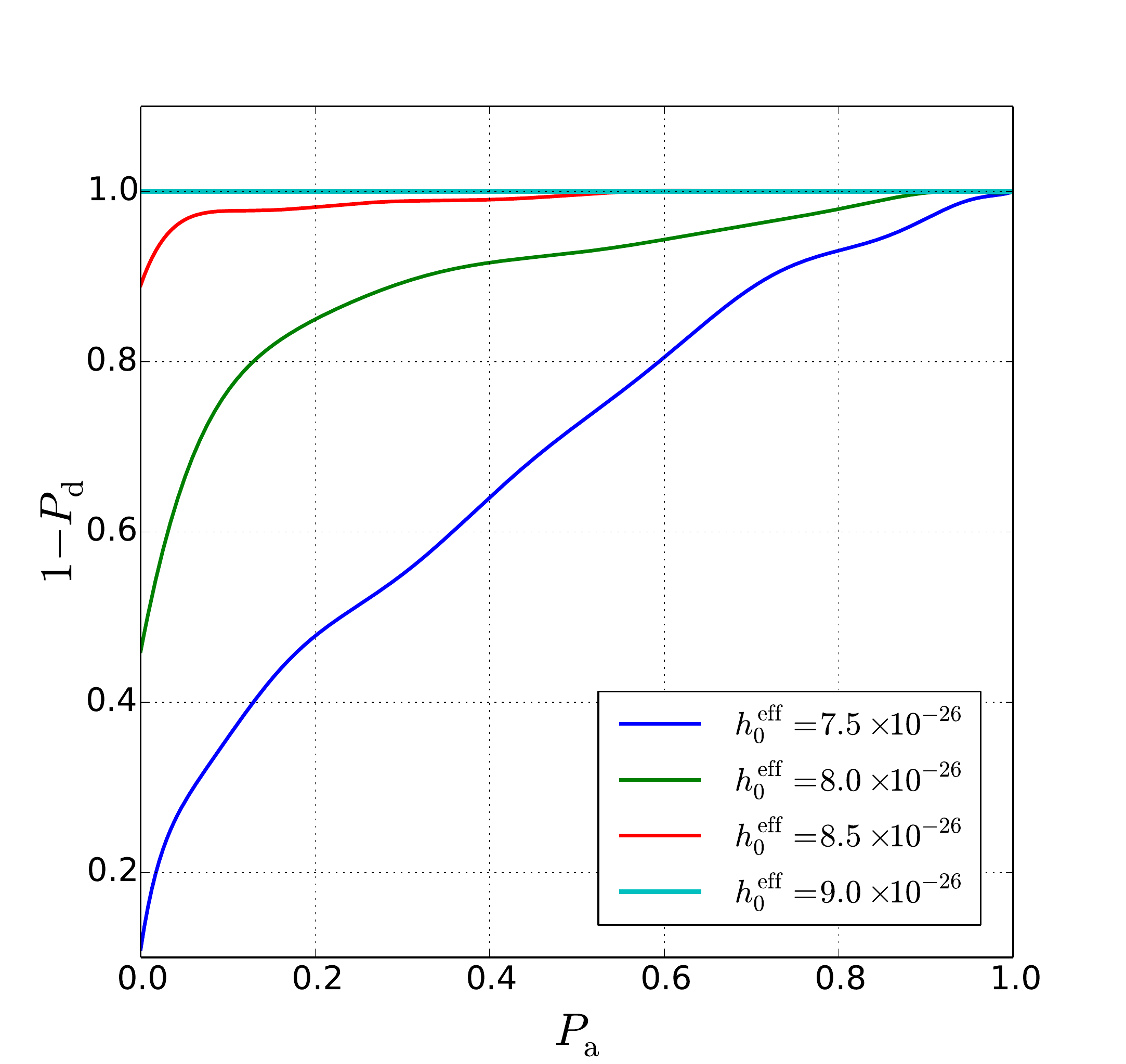}}
	\caption{Receiver operator characteristic (ROC) curves for the searches described in Section \ref{sec:1D-young}. The four curves (from top to bottom) correspond to the four representative wave strains $h_0^{\rm eff}/10^{-26} = 9.0$, 8.5, 8.0, and 7.5. The horizontal and vertical axes indicate the false alarm probability $P_{\rm a}$ and detection probability $1-P_{\rm d}$, respectively. Each curve is based on 200 realizations. Parameters: $T_{\rm drift} =1$\,hr, $N_T = 2000$, $S_h (f)^{1/2} = 4 \times 10^{-24}$\,Hz$^{-1/2}$.}
	\label{fig:roc2}
\end{figure}

Figure \ref{fig:roc2} shows the ROC curves for the searches in Section \ref{sec:1D-young} with four $h_0^{\rm eff}$ values, ranging from $7.5 \times 10^{-26}$ to $9.0 \times 10^{-26}$. The properties of the curves are similar to Figure \ref{fig:roc1}. However, the overall sensitivity degrades by a factor of $\approx 3.5$, with $h_0^{\rm eff,95\%} \approx 8.5 \times 10^{-26}$ ($T_{\rm obs} = 83.3$\,d).

\section{HMM Tracking of $f_0$ and $\dot{f_0}$}
\label{sec:hmm_fdot}
In Sections \ref{sec:hmm_f0} and \ref{sec:simulations_and_sensi}, we show that one-dimensional HMM tracking can be applied to search for any young objects, but the sensitivity degrades when the spin-down rate is too high, e.g., $|\dot{f_0}|\gtrsim 10^{-8}$\,Hz\,s$^{-1}$ and $T_{\rm drift}\lesssim$ a few hours. In this section, we describe a more costly alternative to $f_0$ tracking, which allows relatively longer $T_{\rm drift}$ when the spin-down rate is high. We formulate the tracker as a two-dimensional HMM with hidden state ($f_0, \dot{f_0}$) in Section \ref{sec:f_fdot_tracking} and present simulation examples in Section \ref{sec:2D-simulation}.

\subsection{Transition and emission probabilities}
\label{sec:f_fdot_tracking}
In this implementation, we define a two-dimensional hidden state variable $q(t)=[f_0(t),\dot{f_0}(t)]$ and track $f_0$ and $\dot{f_0}$ jointly. The state variable can take $N_Q=N_{f_0}N_{\dot{f_0}}$ possible discrete values $q_{ij} \in\{q_{11}, \cdots, q_{N_{f_0}N_{\dot{f_0}}}\}$, where $i$ and $j$ index $f_0$ and $\dot{f_0}$ bins, respectively, and $N_{f_0}$ and $N_{\dot{f_0}}$ are the total number of $f_0$ and $\dot{f_0}$ bins, respectively. 

The discrete hidden states are mapped one-to-one to the two-dimensional array of bins in the output of the estimator $\mathcal{F}(f_0,\dot{f_0})$ computed over $T_{\rm drift}$.\footnote{The $\mathcal{F}$-statistic is computed as a function of $f_0(t)$ and $\dot{f_0}(t)$ at a given reference time. We normally choose the start time $t_n$ of the interval as the reference time.} The $f_0$ and $\dot{f_0}$ bin sizes $\Delta f_0$ and $\Delta \dot{f_0}$ are selected using a phase metric described in Appendix \ref{sec:cost}. Assuming that the spin-down evolution of a neutron star is smooth (i.e. no glitches) and that $\ddot{f_0}(t)$ is bounded, we can always choose an intermediate time-scale $T_{\rm drift}$ for a particular astrophysical source, $T_{\rm SFT} < T_{\rm drift} < T_{\rm obs}$, to satisfy 
\begin{equation}
\label{eqn:int_T_drift_2}
\left|\int_t^{t+T_{\rm drift}}dt' \ddot{f_0}(t')\right| < \Delta \dot{f_0}
\end{equation}
for $0<t<T_{\rm obs}$. We calculate $f_0(t_{n+1})$ from the estimated $f_0(t_n)$ and $\dot{f_0}(t_n)$ according to\footnote{Alternatively, if we track $f_0(t)$ and $\dot{f_0}(t)$ independently, another constraint on $T_{\rm drift}$ is imposed by $\Delta f_0$, given by (\ref{eqn:int_T_drift}). In other words, we cannot use longer $T_{\rm drift}$ than that in the $f_0$ tracking. Hence we do not track $f_0$ and $\dot{f_0}$ independently and choose $T_{\rm drift}$ to satisfy (\ref{eqn:int_T_drift_2}) only.}
\begin{equation}
\label{eqn:next_f0}
f_0(t_{n+1}) = f_0(t_n) + \dot{f_0} (t_n)T_{\rm drift}.
\end{equation}

If we update $f_0(t_{n+1})$ according to (\ref{eqn:next_f0}), the transition probability matrix becomes 
\begin{eqnarray}
\nonumber A_{q_{i-\Delta i,  j+1} q_{ij}} &=& A_{q_{i-\Delta i,  j} q_{ij}} \\
\label{eqn:trans_matrix_2}
&=& \{2[\Delta i_{\rm max}(j) - \Delta i_{\rm min}(j)+1]\}^{-1},
\end{eqnarray}
with all other terms being zero. In (\ref{eqn:trans_matrix_2}), $\Delta i$ takes integer values $\Delta i_{\rm min}(j) \leq \Delta i \leq \Delta i_{\rm max}(j)$ with
\begin{eqnarray}
\Delta i_{\rm min}(j) &=& { \rm floor}(|\dot{f_0}_{j+1}|T_{\rm drift}/\Delta f_0), \\
\Delta i_{\rm max}(j) &=& {\rm ceil}(|\dot{f_0}_j|T_{\rm drift}/\Delta f_0),
\end{eqnarray}
where ${\rm floor}(x)$ denotes the largest integer smaller than or equal to $x$, ${\rm ceil}(x)$ denotes the smallest integer larger than or equal to $x$, and $\dot{f_0}_j$ is the value of $\dot{f_0}$ in the $j$-th $\dot{f_0}$ bin. The detailed derivation of (\ref{eqn:trans_matrix_2}) is given in Appendix \ref{sec:transition_matrix_proof}.

The emission probability is given by
\begin{eqnarray}
\label{eqn:emi_prob_matrix_2}
\nonumber
L_{o(t) q_{ij}} &=& P [o(t)|{f_0}_i \leq f_0(t) \leq {f_0}_i+\Delta f_0, \\
&&\dot{f_0}_j \leq \dot{f_0}(t) \leq \dot{f_0}_j+\Delta \dot{f_0}]\\ 
\label{eqn:matrix_propto_2}
&\propto& \exp[\mathcal{F}({f_0}_i,\dot{f_0}_j)].
\end{eqnarray}
We choose a uniform prior in both $f_0$ and $\dot{f_0}$, viz.
\begin{eqnarray}
\label{eqn:prior_2}
\Pi_{q_{ij}} = N_Q^{-1}.
\end{eqnarray}

\subsection{Abridged mock search}
\label{sec:2D-simulation}
In this section we demonstrate the ($f_0, \dot{f_0}$) HMM tracker using synthetic data. To make a fair comparison with the $f_0$ tracker, we conduct an abridged version of a mock search for the rapidly spinning down signal simulated in Section \ref{sec:1D-young}, with the same injection parameters as in Tables \ref{tab:inj-1D-slow} and \ref{tab:inj-fast-sd}. We choose $T_{\rm drift} = 50$\,hr ($N_T=40$) to satisfy (\ref{eqn:int_T_drift_2}) and use the search parameters in Table \ref{tab:2D-params}. The $\mathcal{F}$-statistic is computed over a 1-Hz frequency band as a function of $f_0$ and $\dot{f_0}$ for each segment. For demonstration purposes, only five values of $\dot{f_0}$ are searched in a range containing the injected $\dot{f_0}$ to save time, i.e., the phase metric is not computed. 

The results are presented in Table \ref{tab:2d-results}. Compared to the performance displayed in Table \ref{tab:results-1D-fast} using $f_0$ tracking, the ($f_0, \dot{f_0}$) tracking can detect a signal about three times weaker. We calculate the RMSE $\varepsilon_{f_0}$ in $f_0$ between the optimal Viterbi path and the injected signal (in Hz and in units of $\Delta f_0$). We do the same for the RMSE $\varepsilon_{\dot{f_0}}$ in $\dot{f_0}$ (in Hz\,s$^{-1}$ and in units of $\Delta \dot{f_0}$). In a real search, we consider candidates for follow-up and further scrutiny, if $S$ exceeds a threshold $S_{\rm th}$ set by the desired false alarm and false dismissal probabilities, as shown in Section \ref{sec:simulations_and_sensi}. The value of $S_{\rm th}$ depends strongly on $N_Q$ and hence the two-dimensional ($f_0,\dot{f_0}$) parameter space. Systematic Monte-Carlo simulations are required in practice to calculate $S_{\rm th}$ for each HMM implementation, an exercise lying outside the scope of this paper. Instead, in this section, we adopt the following rule of thumb: the injected signal is deemed to be detected if we obtain $\varepsilon_{\dot{f_0}} < 0.5 \Delta \dot{f_0}$ and $\varepsilon_{f_0} < 10 \Delta f_0$. The errors are introduced mostly because HMM takes discrete values of $f_0$ and $\dot{f_0}$, while the injected signal $f_0(t)$ and $\dot{f_0}(t)$ can be any value within a bin. Since we calculate $f_0(t_{n+1})$ from the estimated $f_0(t_n)$ and $\dot{f_0}(t_n)$, $\varepsilon_{f_0}$ accumulates to a few $\Delta f_0$ after $N_T$ steps, introduced by $\varepsilon_{\dot{f_0}}$. 

\begin{figure*}
	\centering
	\subfigure[]
	{
		\label{fig:2D-4e-26}
		\scalebox{0.32}{\includegraphics{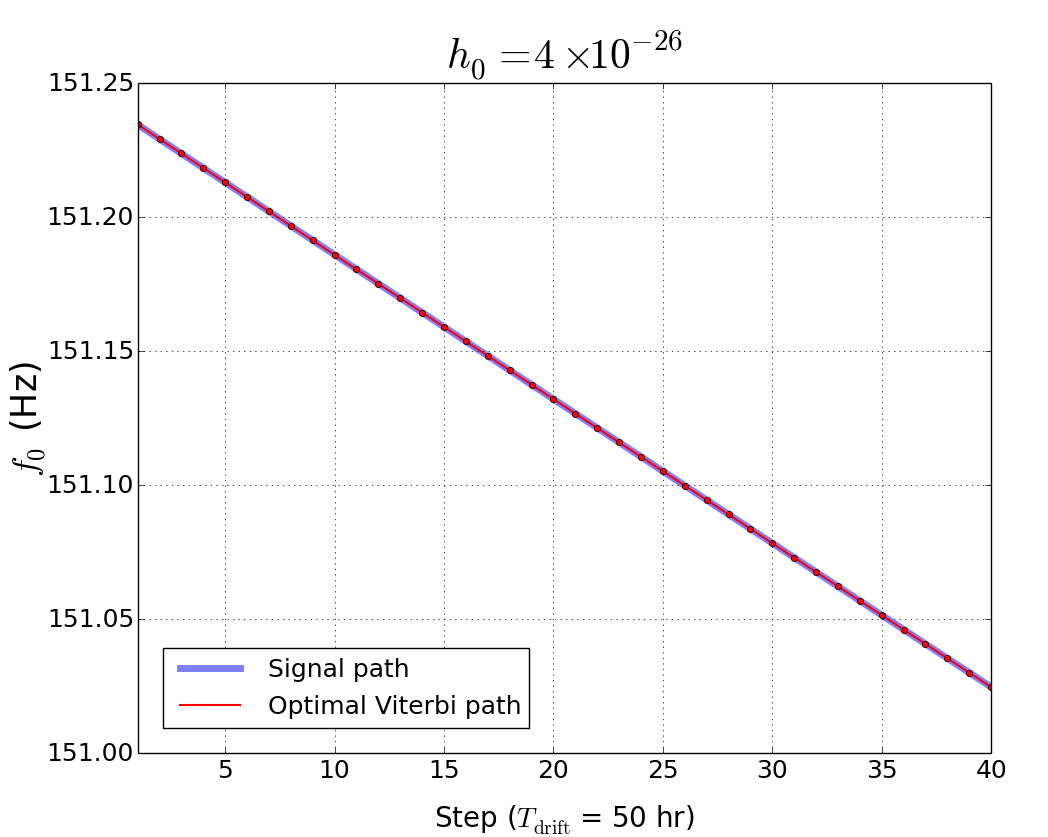}}
		\scalebox{0.32}{\includegraphics{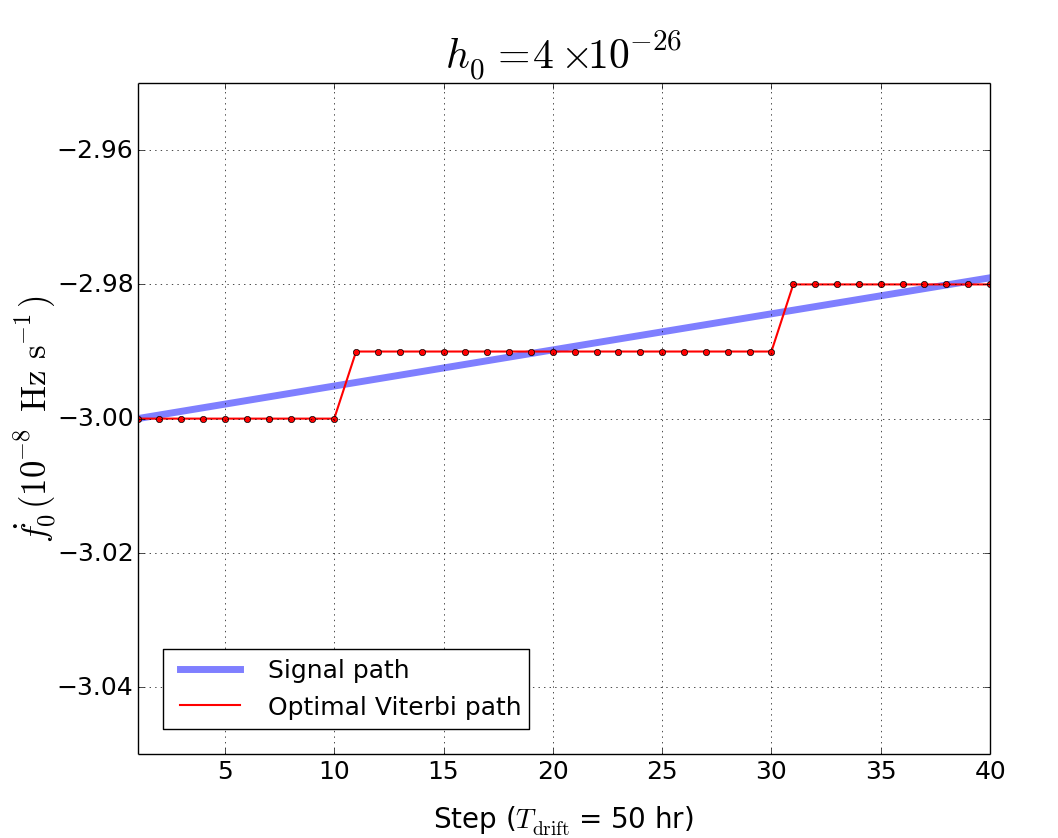}}
	}
	\subfigure[]
	{
		\label{fig:2D-3e-26}
		\scalebox{0.32}{\includegraphics{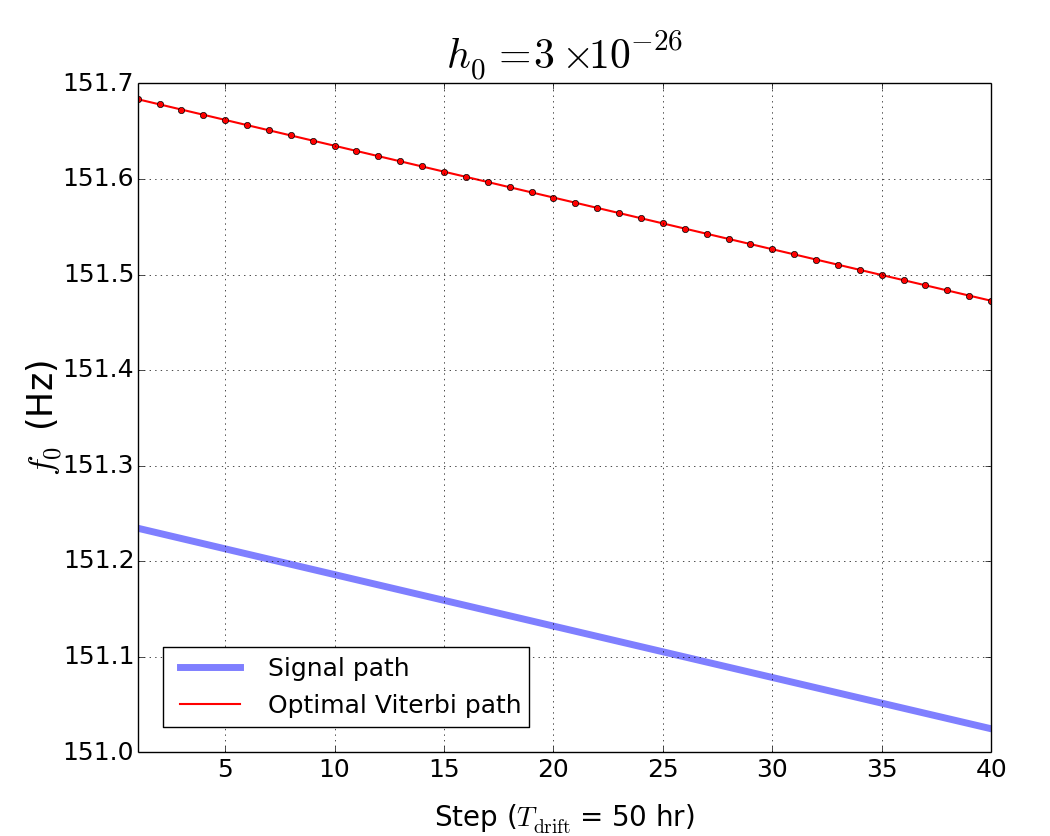}}
		\scalebox{0.32}{\includegraphics{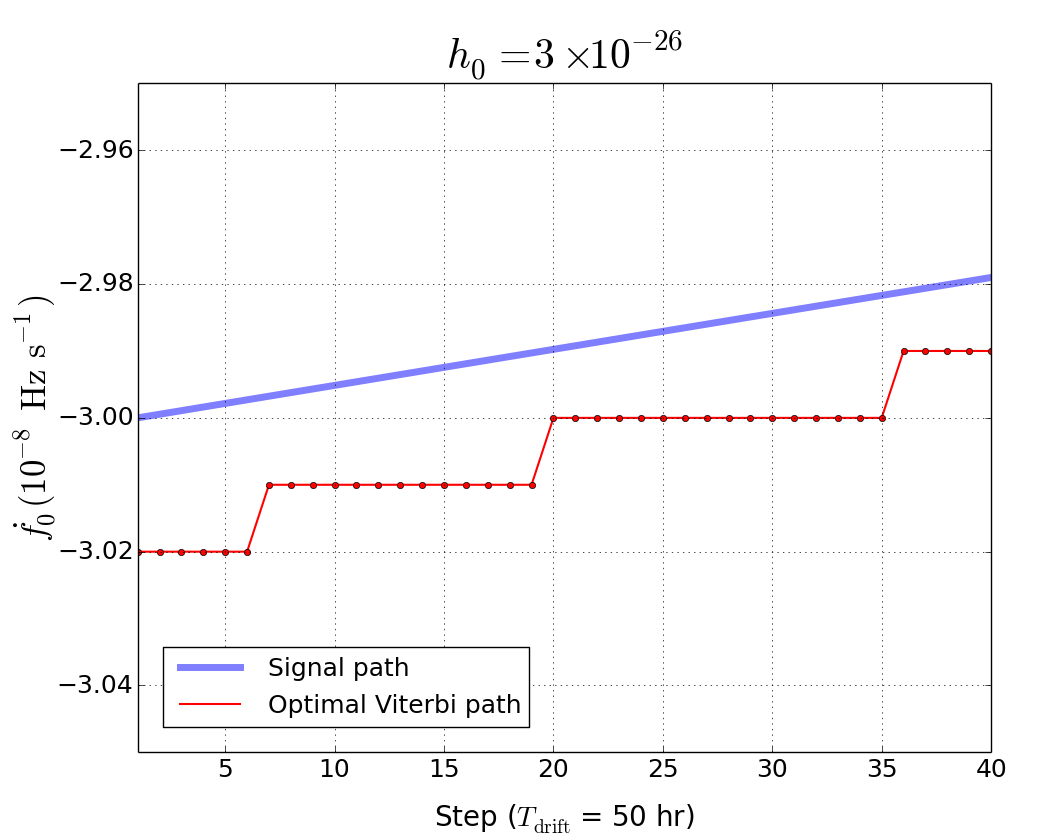}}
	}
	\caption{Injected $f_0(t)$ and $\dot{f_0}(t)$ (blue curves) and optimal Viterbi paths (red curves) for the last two injected signals in Table \ref{tab:2d-results} with (a) $h_0= 4 \times 10^{-26}$ and (b) $h_0 = 3 \times 10^{-26}$. The top left and top right panels show good matches for both $f_0$ and $\dot{f_0}$, respectively. The red curve in the left panel fluctuates around the blue curve with $\varepsilon_{f_0} = 1.80 \times 10^{-5}$\,Hz, which is too small to be seen in the plot ($\varepsilon_{f_0} \ll \dot{f_0}T_{\rm obs} \approx 0.2$\,Hz). In the lower two panels, the injected signal is not detected. The horizontal axes are in units of HMM steps with $T_{\rm drift} = 50$\,hr for each step ($N_T = 40$).}
	\label{fig:2D_tracking_results}
\end{figure*}

Figure \ref{fig:2D_tracking_results} displays the optimal Viterbi paths (red curves) and the true paths $f_0(t)$ and $\dot{f_0}(t)$ (blue curves) for the two weakest injections (a) $h_0= 4 \times 10^{-26}$ and (b) $h_0 = 3 \times 10^{-26}$. The left and right panels show $f_0$ and $\dot{f_0}$, respectively. In Figure \ref{fig:2D-4e-26}, the optimal Viterbi paths agree with $f_0(t)$ and $\dot{f_0}(t)$ closely. In the right panel, it is shown that the estimated $\dot{f_0}$ fluctuates within one bin around the injected $\dot{f_0}(t)$. The fluctuations around $f_0$ cannot be seen clearly in the left panel, because $\varepsilon_{f_0} = 1.80 \times 10^{-5}$\,Hz is much smaller than the total change of $f_0$ over $T_{\rm obs}$ ($\approx 0.2$\,Hz). In contrast, Figure \ref{fig:2D-3e-26} shows that the optimal Viterbi paths do not match the injected $f_0(t)$ and $\dot{f_0}(t)$, i.e., the injected signal is not detected.

\begin{table}
	\centering
	\setlength{\tabcolsep}{4pt}
	\renewcommand\arraystretch{1.4}
	\begin{tabular}{lll}
		\hline
		\hline
		Parameter & Value & Unit\\
		\hline
		$f_0$ &151--152 & Hz \\
		$-\dot{f_0}$ & $2.98\times 10^{-8}$--$3.02 \times 10^{-8}$& Hz\,s$^{-1}$ \\
		$T_{\rm drift}$ & 50 & hr \\
		$\Delta f_0$ & $2.78 \times 10^{-6}$ & Hz \\
		$\Delta \dot{f_0}$ & $1 \times 10^{-10}$ & Hz\,s$^{-1}$ \\
		$T_{\rm obs}$ & 83.3 & d\\
		$N_T$ & 40 & --\\
		\hline
		\hline
	\end{tabular}
	\caption[parameters]{Search parameters for the synthetic signals with injection parameters quoted in Tables \ref{tab:inj-1D-slow} and \ref{tab:inj-fast-sd}.}
	\label{tab:2D-params}
\end{table}

\begin{table*}
	\centering
	\setlength{\tabcolsep}{6pt}
	\renewcommand\arraystretch{1.4}
	\begin{tabular}{lllllll}
		\hline
		\hline
		$h_0$ $(10^{-26})$ & Detect? & $S$ & $\varepsilon_{f_0}$ (Hz) & $\varepsilon_{f_0}/\Delta f_0$& $\varepsilon_{\dot{f_0}}$ (Hz\,s$^{-1}$) & $\varepsilon_{\dot{f_0}}/\Delta \dot{f_0}$\\
		\hline
		$8.0$ & $\checkmark$ & 9.0 & $2.89\times 10^{-6}$ & 1.0 & $2.97\times 10^{-11}$& 0.30\\
		$5.0$ & $\checkmark$  & 2.9 & $6.45\times 10^{-6}$ & 2.3 & $3.56\times 10^{-11}$& 0.36\\
		$4.0$ & $\checkmark$  &2.0& $1.80\times 10^{-5}$& 6.5 & $2.85\times 10^{-11}$& 0.29 \\
		$3.0$ & $\times$          & 1.2 & 0.45 & $1.6\times 10^5$ & $1.659\times 10^{-10}$& 1.59\\
		\hline
		\hline
	\end{tabular}
	\caption[]{Results of ($f_0$, $\dot{f_0}$) tracking for injected signals with the parameters in Tables \ref{tab:inj-1D-slow} and \ref{tab:inj-fast-sd}, $T_\text{obs}=83.3$\,d, $T_\text{drift}=50$\,hr, and wave strain $h_0$. The RMSE $\varepsilon_{f_0}$ between the frequency of the optimal Viterbi path and the injected $f_0(t)$ is quoted in Hz and in units of $\Delta f_0 = 2.78 \times 10^{-6}$\,Hz. The RMSE $\varepsilon_{\dot{f_0}}$ between the frequency derivative of the optimal Viterbi path and the injected $\dot{f_0}(t)$ is quoted in Hz\,s$^{-1}$ and in units of $\Delta \dot{f_0} = 1 \times 10^{-10}$\,Hz\,s$^{-1}$. The third column quotes the Viterbi score $S$.}
	\label{tab:2d-results}
\end{table*}

\section{Discussion}
\label{sec:discussion}

\subsection{Cost-sensitivity trade-off}
\label{sec:tradeoff}

In this section, we start by comparing the HMM tracking method to existing stack-slide-based semi-coherent methods and then discuss the cost-sensitivity trade-off between ($f_0, \dot{f_0}$) tracking and $f_0$ tracking. Analytic approximations for the computing cost and sensitivity are described briefly in Appendices \ref{sec:cost} and \ref{sec:sensitivity}.

HMM tracking incoherently combines the $\mathcal{F}$-statistic outputs from $N_T=T_{\rm obs}/T_{\rm drift}$ blocks of data. The computing cost is composed of two parts: (1) calculating the coherent $\mathcal{F}$-statistic (i.e. $L_{o_jq_i}$) for all $N_T$ segments; and (2) recursively maximizing $P(Q|O)$, i.e. solving the HMM. Assuming we use data from two interferometers and search up to the maximum frequency ${f_0}_{\rm max}$, the computing costs of calculating $\mathcal{F}(f_0)$ and $\mathcal{F}(f_0,\dot{f_0})$ over one block of coherent segment $T_{\rm drift}$ are given by
\begin{equation}
	\label{eqn:compute_time_1D}
	\mathcal{T}_{f_0} = 0.46\,{\rm d}\left(\frac{{f_0}_{\rm max}}{0.6\,{\rm kHz}}\right) \left(\frac{T_{\rm drift}}{10\,{\rm d}}\right)^2 \left(\frac{1}{N_{\rm core}}\right),
\end{equation}
and
\begin{equation}
\label{eqn:compute_time_2D}
\mathcal{T}_{f_0,\dot{f_0}} = 0.36\,{\rm d} \left(\frac{{f_0}_{\rm max}}{0.6\,{\rm kHz}}\right)^2\left(\frac{0.3\,{\rm kyr}}{\tau}\right)\left(\frac{T_{\rm drift}}{10\,{\rm d}}\right)^4\left(\frac{10^3}{N_{\rm core}}\right),
\end{equation}
respectively, where $N_{\rm core}$ is the number of cores running in parallel (see details in Appendix \ref{sec:cost}). The Viterbi algorithm computes $Q^*(O)$ via $(N_T+1)N_Q \ln N_Q$ operations \cite{Suvorova2016}. For example, in a 1-Hz sub-band with $N_Q = 2\times 10^6$ and $N_T = 36$, it takes $\lesssim 30$\,s to compute $Q^*(O)$ but $\gtrsim 1$\,hr to compute $N_T$ blocks of the $\mathcal{F}$-statistic. Hence the total computing cost is dominated by the cost of computing $N_T$ blocks of the $\mathcal{F}$-statistic, scaling as $\mathcal{T}_{f_0} \propto N_T {f_0}_{\rm max} T_{\rm drift}^2$ for $f_0$ tracking, and $\mathcal{T}_{f_0,\dot{f_0}}\propto N_T {f_0}_{\rm max}^2 \tau^{-1} T_{\rm drift}^4$ for ($f_0, \dot{f_0}$) tracking. 

Compared to a fully coherent $\mathcal{F}$-statistic search, the cost saving conferred by the HMM tracker is similar to other $\mathcal{F}$-statistic-based semi-coherent methods, when only $f_0$ or ($f_0, \dot{f_0}$) needs to be searched. Theoretically, the sensitivity of the HMM tracker is also comparable to other $\mathcal{F}$-statistic-based semi-coherent searches. Hence the HMM tracker performs on par with other semi-coherent methods, as long as the spin-down rate is moderate, except that it is more robust against timing noise, as demonstrated in Section \ref{sec:1D-simulation-sw}.

\begin{figure}
	\centering
	\scalebox{0.35}{\includegraphics{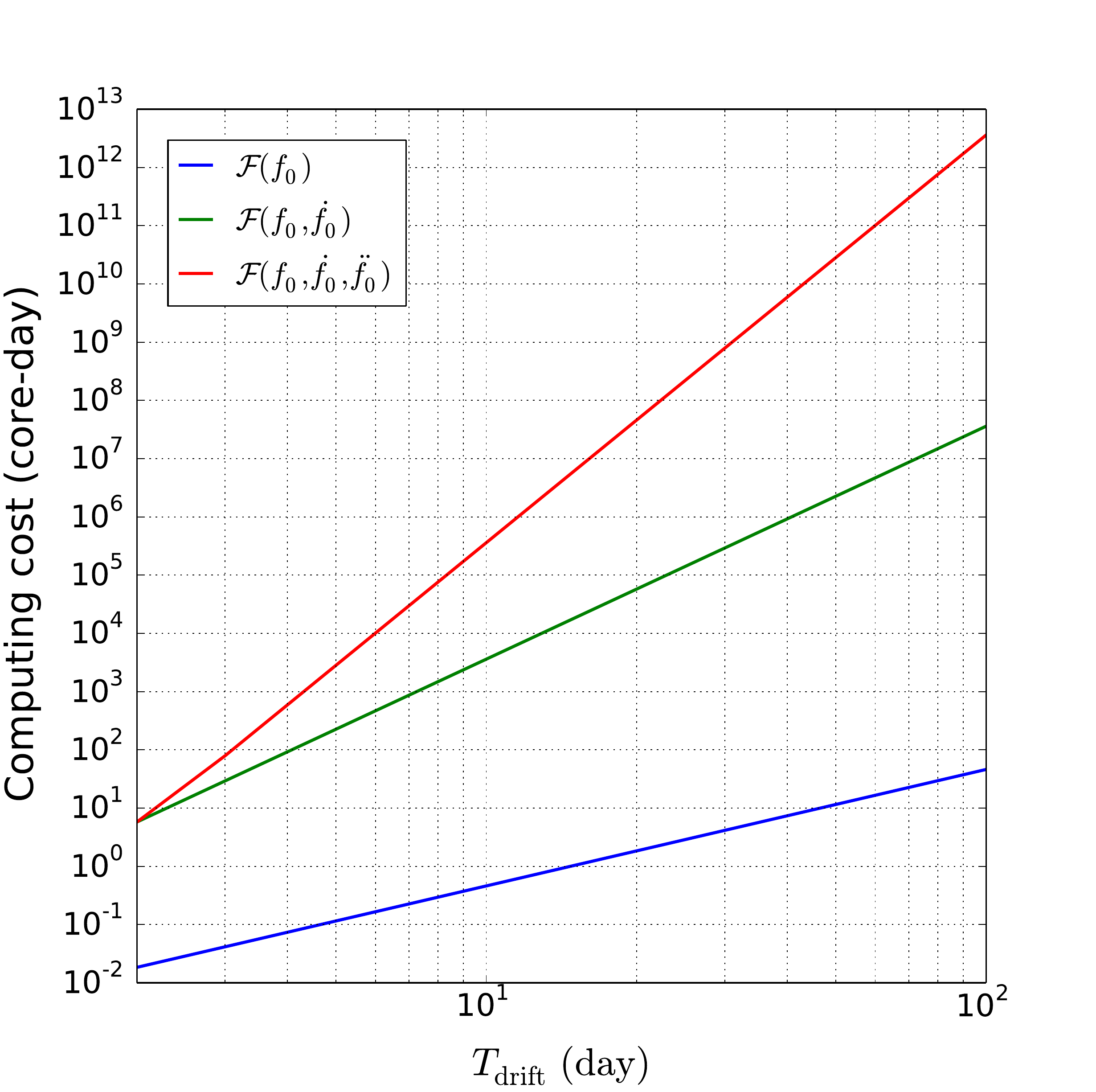}}
	\caption{Computing cost (in core-days) of a coherent $\mathcal{F}$-statistic search as a function of $T_{\rm drift}$ (in days). The three curves (from bottom to top) represent searching $f_0$ only, ($f_0, \dot{f_0}$), and ($f_0, \dot{f_0}, \ddot{f_0}$), respectively. Parameters: ${f_0}_{\rm max}=600$\,Hz, $\tau = 0.03$\,kyr.}
	\label{fig:fstat-cost}
\end{figure}

When higher-order derivatives of the frequency are required to be searched for very young objects (e.g., $\tau \lesssim 0.03$\,kyr), e.g. in a stack-slide search, the cost of computing $\mathcal{F}$-statistic grows geometrically as $T_{\rm drift}$ increases. Figure~\ref{fig:fstat-cost} shows the cost of calculating the $\mathcal{F}$-statistic over one block of duration $T_{\rm drift}$ for a target with  $\tau = 0.03$\,kyr up to ${f_0}_{\rm max}=600$\,Hz. The three curves, from bottom to top, represent calculating $\mathcal{F}(f_0)$, $\mathcal{F}(f_0,\dot{f_0})$, and $\mathcal{F}(f_0, \dot{f_0}, \ddot{f_0})$, respectively. For example, it requires $\sim 10^3$ core-day to compute $\mathcal{F}(f_0, \dot{f_0}, \ddot{f_0})$ for a single block of duration $T_{\rm drift}=4$\,d. When $\dddot{f_0}$ needs to be considered, the cost of calculating $\mathcal{F}(f_0, \dot{f_0}, \ddot{f_0}, \dddot{f_0})$ becomes prohibitive even for $T_{\rm drift}<1$\,d. Under these circumstances, the HMM tracker comes into its own; it allows an efficient search for rapidly evolving signals without searching high-order frequency derivatives.

Given fixed $T_{\rm obs}$, one can tune $T_{\rm drift}$ to trade off sensitivity against computing cost for a particular target. Table \ref{tab:trade-off-scaling} shows the theoretical scalings of sensitivity and cost as a function of $T_{\rm drift}$ for the two HMM implementations described in Section \ref{sec:hmm_f0} and \ref{sec:hmm_fdot}. In reality, the scalings vary with many factors, including $N_T$, $P_{\rm a}$, $P_{\rm d}$, and the noise statistics, as discussed in detail in Ref. \cite{Wette2012,Prix2012}. In this paper, we include the theoretical scalings to allow quick order-of-magnitude comparisons, but we emphasize that they are not a substitute for Monte-Carlo simulations. The sensitivities of $f_0$ tracking and ($f_0, \dot{f_0}$) tracking scale the same way with $T_{\rm drift}$. An $(f_0, \dot{f_0})$ search allows longer $T_{\rm drift}$ and hence in practice is always more sensitive than an $f_0$ search. However, an $f_0$ search is always faster.

\begin{figure}
	\centering
	\scalebox{0.3}{\includegraphics{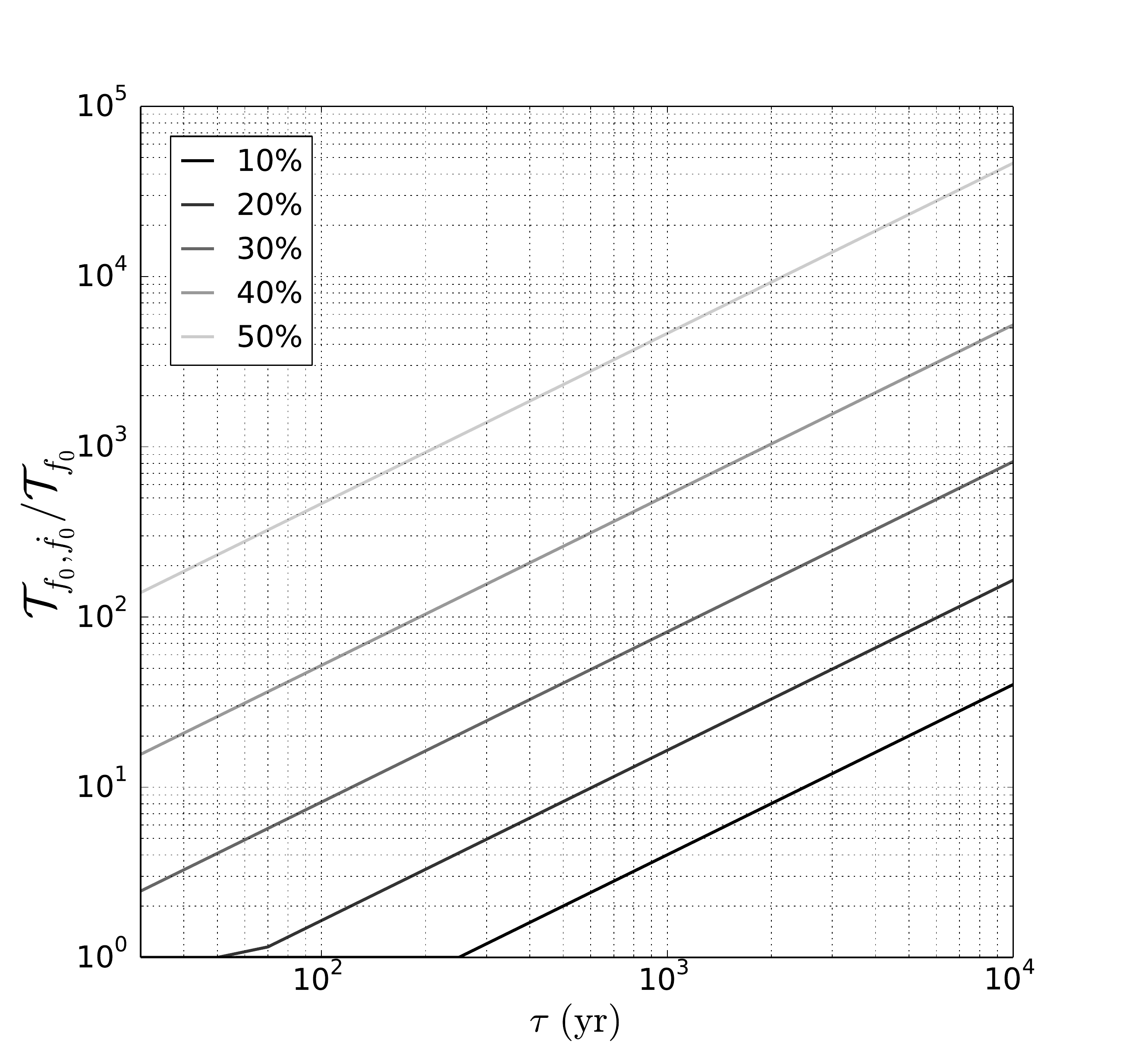}}
	\caption{Ratio of the ($f_0,\dot{f_0}$) tracking cost $\mathcal{T}_{f_0,\dot{f_0}}$ divided by the $f_0$ tracking cost $\mathcal{T}_{f_0}$ as a function of the target age $\tau$, required to improve the sensitivity by 10\%--50\% (from bottom curve to top curve) by switching from $f_0$ tracking to ($f_0,\dot{f_0}$) tracking. The $f_0$ tracking is always faster but less sensitive than ($f_0,\dot{f_0}$) tracking. For a given percentage of sensitivity improvement, the cost required for choosing ($f_0,\dot{f_0}$) tracking rather than $f_0$ tracking increases with $\tau$.}
	\label{fig:trade-off}
\end{figure}

Figure \ref{fig:trade-off} plots the ratio $\mathcal{T}_{f_0,\dot{f_0}}/\mathcal{T}_{f_0}$ as a function of $\tau$. The curves, from bottom to top, represent achieving 10\%--50\% better sensitivity by switching from $f_0$ tracking to ($f_0,\dot{f_0}$) tracking. We can always achieve better sensitivity using ($f_0,\dot{f_0}$) tracking compared to $f_0$ tracking. However, $\mathcal{T}_{f_0,\dot{f_0}}/\mathcal{T}_{f_0}$ is approximately proportional to $\tau$ and increases exponentially with the percentage of sensitivity improvement. 

\begin{table}
	\centering
	\setlength{\tabcolsep}{8pt}
	\renewcommand\arraystretch{1.5}
	\begin{tabular}{lll}
		\hline
		\hline
		Tracking & Sensitivity & Cost\\
		\hline
		$f_0$ & $T_{\rm drift}^{-1/4}$& $T_{\rm drift}$\\
		$f_0$ and $\dot{f_0}$ & $T_{\rm drift}^{-1/4}$ & $T_{\rm drift}^3$\\
		\hline
		\hline
	\end{tabular}
	\caption{Theoretical scalings of sensitivity and computing cost with drift time-scale $T_{\rm drift}$ for $f_0$ tracking (Section \ref{sec:hmm_f0}) and $(f_0, \dot{f_0})$ tracking (Section \ref{sec:hmm_fdot}).}
	\label{tab:trade-off-scaling}
\end{table}

\subsection{Spin down of young objects with age $\lesssim {f_0}_{\rm birth}/\dot{f_0}_{\rm birth}$}
\label{sec:spindown_vs_age}
The true age of a young neutron star may be significantly less than its characteristic spin-down time-scale at birth, ${f_0}_{\rm birth}/[(n-1){\dot{f_0}}_{\rm birth}]$, depending on its ellipticity and magnetization. To investigate this scenario, we approximate the braking law with a power law in the usual way, viz. $\dot{f_0}(t) = -\eta f_0(t)^n$, with $\eta \propto B_0^2$ and $2\lesssim n <3$ if the torque is electromagnetically dominated, and $\eta \propto \epsilon^2$ and $n=5$ if the torque is dominated by gravitational radiation reaction, where $\epsilon$ is the ellipticity. Integrating the braking law with respect to $t$, we find that the characteristic spin-down time-scale of the signal is given by \cite{Wette2008,Sun2016}
\begin{equation}
\label{eq:spindown-ts}
-f_0/\dot{f_0}=\left|\xi\right|^{-1}\tau
\end{equation}
with
\begin{equation}
\label{eq:xi}
\xi = \frac{1}{n-1}\left[1-\left(\frac{{f_0}_{\rm birth}}{f_0}\right)^{1-n}\right].
\end{equation}
The term $({f_0}_{\rm birth}/{f_0})^{1-n}$ is normally neglected under the assumption $f_0 \ll {f_0}_{\rm birth}$, yielding $\dot{f_0} \approx -f_0(n-1)^{-1} \tau^{-1}$\cite{Abadie2010,Chung2011}. However, this assumption does not necessarily apply to young objects (e.g. $\tau = 0.03$\,kyr for SNR 1987A), for which we obtain $\left|\xi \right| \lesssim 0.05$ for ${f_0}_{\rm birth} \lesssim 600$\,Hz and $B_0 \lesssim 6\times 10^{12}$\,G with $n=3$. A detailed discussion can be found in Section IIB of Ref. \cite{Sun2016}. 

The indirect upper limit on $h_0$ derived from energy conservation is given by \cite{Wette2008,Chung2011,Riles2013}
\begin{equation}
\label{eq:h0upperlim}
h_0\leq \frac{1}{D}\left(\frac{5GI\left|\xi\right|}{2c^3\tau}\right)^{1/2},
\end{equation}
where $G$ is Newton's gravitational constant, $c$ is the speed of light, and $D$ is the distance to the source. The indirect limit on $h_0$ is lowered because of the second term in (\ref{eq:xi}). On the other hand, the slower spin-down rate $\dot{f_0}= -f_0 |\xi|\tau^{-1}$ benefits HMM tracking by allowing longer $T_{\rm drift}$. If we consider a young object with $\tau = 0.03$\,kyr as an example, the impact of having $|\xi|=0.05$ translates into raising $T_{\rm drift}$ by a factor of $\approx 3$.

\section{Conclusion}
\label{sec:conclusion}

In this paper, we describe two practical implementations of an efficient HMM tracker, combined with the maximum likelihood matched filter $\mathcal{F}$-statistic, to economically search for continuous gravitational waves from young neutron stars in SNRs. The HMM incoherently combines the coherent $\mathcal{F}$-statistic outputs from multiple ($N_T$) data blocks of duration $T_{\rm drift}$. It tracks rapid, secular spin down without searching high-order derivatives of the signal frequency. The first implementation, tracking $f_0$ alone, can simultaneously surmount two challenges in young SNR searches: rapid spin down and stochastic timing noise. Three scenarios for different spin-down and timing-noise time-scales are discussed. Given $T_{\rm obs} = 83.3$\,d, we obtain $h_0^{\rm eff,95\%} \approx 2.4 \times 10^{-26}$ for both weak and strong timing noise in the first two scenarios ($\tau \gtrsim 5$\,kyr) and $h_0^{\rm eff,95\%} \approx 8.5 \times 10^{-26}$ in the last scenario ($\tau \lesssim 0.03$\,kyr), with $P_{\rm a}=1\%$. We expect that $h_0^{95\%}$ is more conservative than the quoted $h_0^{\rm eff,95\%}$ for unknown $\cos\iota$ based on scaling given by (\ref{eqn:h0eff}). The second implementation, tracking $f_0$ and $\dot{f_0}$, allows longer $T_{\rm drift}$ and hence improves the sensitivity by a factor of a few. The first implementation is always faster and more robust against timing noise. One can achieve better sensitivity by switching from the first implementation to the second. However, it increases the computing cost by two to three orders of magnitude, depending on $\tau$. 

An optimized $\mathcal{F}$-statistic-based semi-coherent Einstein@Home search for Cas A (${f_0}_{\rm max}=1.5$\,kHz) in the Advanced LIGO O1 run costs approximately $2.7\times 10^5$\,core-day, yielding 90\% confidence strain upper limit $1.4 \times 10^{-25}$ \cite{Ming2017}. Assuming the same parameters, the method discussed in this paper is expected to provide comparable sensitivity but cost $\sim 10^4$\,core-day. The advantage of HMM tracking grows in searches for younger targets, e.g., SNR 1987A.

The methods described in this paper can be applied to extending the searches for the SNRs listed in Ref. \cite{Aasi2015-snr}, which are restricted to coherent segments of duration $T_{\rm drift}\leq 25.3$\,d, using the new data from a whole Advanced LIGO observing run. In addition, the recent work by \citet{Anderson2017} has identified 76 new Galactic SNR candidates, some of which may be promising candidates for gravitational-wave sources, if the SNR associates are confirmed. The $f_0$ tracker can be applied to search for targets that are poorly modelled, e.g., a long transient post-merger signal from the binary neutron star merger GW170817 \cite{postmerger2017} with spin-down time-scale $\sim 10^2$--$10^4$\,s. Some modifications are needed, e.g. $L_{o_jq_i}$ should be calculated from the power in SFT bins rather than the $\mathcal{F}$-statistic, because the Earth's rotation can be neglected.

To carry out a search using the methods presented in this paper, the following steps need to be completed in preparation. First, the search parameter ranges need to be determined systematically. The $f_0$ range is normally chosen to equal the band where the estimated strain sensitivity is below the indirect, $\tau$-based limit [see (\ref{eq:h0upperlim})]. General equations (\ref{eqn:dotf_range}) and (\ref{eqn:ddotf_range}) for calculating $\dot{f_0}$ and $\ddot{f_0}$ are given in Section \ref{sec:search_para_range}. Second, search parameter resolutions need to be calculated using the metric described in Appendix \ref{sec:cost} given a desired mismatch. Third, a systematic Monte-Carlo simulation is required for each implementation to determine the detection threshold $S_{\rm th}$ given false alarm and false dismissal probabilities.

\section{Acknowledgements}
We would like to thank Ra Inta and the LIGO Scientific Collaboration Continuous Wave Working Group for detailed comments and informative discussions. L. Sun is supported by Australia Research Training Program Stipend Scholarship. The research was supported by Australian Research Council (ARC) Discovery Project DP110103347 and the ARC Centre of Excellence for Gravitational Wave Discovery CE17010004. 

\appendix

\section{Viterbi algorithm}
\label{sec:viterbi}
The principle of optimality \cite{Bellman1957} demonstrates that in our special case, all subpaths ${Q^*}^{(k)}$ made up of the first $k$ steps in $Q^*(O)$ are optimal for $1\leq k \leq N_T$. In that sense, the classic Viterbi algorithm \cite{Viterbi1967} provides a recursive, computationally efficient solution to computing $Q^*(O)$ in a HMM, reducing the number of operations from $N_Q^{N_T+1}$ to $(N_T+1)N_Q \ln N_Q$ by binary maximization \cite{Quinn2001}. A full description of the algorithm can be found in Section II D of Ref. \cite{Suvorova2016}. At every forward step $k$ ($1\leq k \leq N_T$) in the recursion, the algorithm eliminates all but $N_Q$ possible state sequences, and stores the $N_Q$ maximum probabilities
\begin{equation}
\delta_{q_i}(t_k) = L_{o(t_k)q_i} \mathop{\max} \limits_{1 \leq j \leq N_Q} [A_{q_i q_j}\delta_{q_j}(t_{k-1})], 
\end{equation}
and previous-step states leading to the retained most likely sequence
\begin{equation}
\Phi_{q_i}(t_k) = \mathop{\arg \max} \limits_{1 \leq j \leq N_Q} [A_{q_i q_j}\delta_{q_j}(t_{k-1})].
\end{equation}
When backtracking, for $0 \leq k \leq N_T -1$, we reconstruct the optimal Viterbi path according to
\begin{equation}
q^*(t_k) = \Phi_{q^*(t_{k+1})}(t_{k+1}).
\end{equation}

\section{Phase metric and computing cost}
\label{sec:cost}
The costs of computing the $\mathcal{F}$-statistic (i.e. $L_{o_jq_i}$) and recursively maximizing $P(Q|O)$ depend on the template spacing. We start by discussing the template spacing and cost for a general $\mathcal{F}$-statistic search. In order to optimize the template spacing, a phase metric is defined. It expresses the signal-to-noise ratio as a function of template spacing along each parameter axis (e.g., $f_0$, $\dot{f_0}$, $\ddot{f_0}$, $\cdots$). The mismatch $m$ is defined as the fractional reduction of $\mathcal{F}$-statistic power caused by discrete parameter sampling, with \cite{Brady1998,Owen1996,Whitbeck2006}
\begin{equation}
\label{eqn:mu}
m = \sum_{i,j}g_{ij}\Delta\lambda^i\Delta\lambda^j,
\end{equation} 
\begin{equation}
\label{eqn:para_metric}
g_{ij} = \frac{4\pi^2(i+1)(j+1)T_{\rm drift}^{i+j+2}}{(i+2)!(j+2)!(i+j+3)}.
\end{equation}
The indices $i$ and $j$ take integer values from 0 to $k$, where $k$ indicates the highest-order frequency derivative considered (e.g. $k=2$ for searching up to $\ddot{f_0}$), and $\Delta\lambda^i$ denotes the offset between the true value and the closest template of the $i$-th parameter. For example, the maximum value of $\Delta\lambda^0$ is half the frequency bin width $\Delta {f_0}$, because the signal frequency falls halfway between two templates in the worst case. We choose to adopt $m \leq 0.2$ in line with the Cas A search in S5 data \cite{Abadie2010} and the SNR searches in S6 data \cite{Aasi2015-snr}. The highest frequency derivative needed is the largest integer $k$ satisfying $g_{kk}[{f_0}^{(k)}_{\rm max} - {f_0}^{(k)}_{\rm min}]^2 >m$ (no summation over $k$ implied), where ${f_0}^{(k)}_{\rm max}$ and ${f_0}^{(k)}_{\rm min}$ are the maximum and minimum $k$-th frequency derivative. In practice, we can choose the bin size of the $i$-th frequency derivative $\Delta {f_0}^{(i)}$ using the diagonal terms of (\ref{eqn:para_metric}) to satisfy 
\begin{equation}
\label{eqn:bin_size}
\sum_{i=0}^k g_{ii}[\Delta {f_0}^{(i)}]^2< 4m.
\end{equation}

Monte-Carlo simulations are needed to accurately calculate the required bin sizes for a given $m$. Taking into consideration the off-diagonal terms of (\ref{eqn:para_metric}) yields bin sizes close to the empirical Monte-Carlo results. A tiling algorithm is described in detail in Ref. \cite{Reinhard2007}. Combining (\ref{eqn:dotf_range}), (\ref{eqn:ddotf_range}) and (\ref{eqn:para_metric}), the number of templates $\mathcal{N}$ needed for $k=2$ is \cite{Reinhard2007,Wette2008}
\begin{equation}
\label{eqn:num_templates}
\mathcal{N} = 0.35m^{-3/2} ({\rm det}\,g)^{1/2}{f_0}_{\rm max}^3\tau^{-3},
\end{equation}
with ${f_0}_{\rm min} \ll {f_0}_{\rm max}$ typically.

The computing time of a coherent $\mathcal{F}$-statistic search over one block of duration $T_{\rm drift}$ is given by  
\begin{equation}
\label{eqn:compute_time}
\mathcal{T} = \kappa \mathcal{N}  \beta N_{\rm ifo}  T_{\rm drift}T_{\rm SFT}^{-1} ,
\end{equation}
where $\kappa$ is the time to compute the $\mathcal{F}$-statistic per template per SFT,\footnote{The value of $\kappa$ depends on $T_{\rm SFT}$ and the CPU architecture. An example in Section~5 of Ref. \cite{Wette2008} quotes $\kappa=6\times 10^{-7}$\,s ($T_{\rm SFT}=1800$\,s) on Australian Partnership for Advanced Computing (APAC) resources. We adopt a more recent estimate, $\kappa=4\times 10^{-8}$\,s, in this paper.}  $N_{\rm ifo}$ is the number of interferometers, and $\beta$ is the percentage of time that the interferometers collect data (i.e., duty cycle). For most of the young targets discussed in Ref. \cite{Aasi2015-snr}, $\ddot{f_0}$ is normally small. Only a few $\ddot{f_0}$ values need to be searched. For example, we obtain $\ddot{f_0} \lesssim 1\times 10^{-18}$\,Hz\,s$^{-2}$ from (\ref{eqn:dotf_range}) and (\ref{eqn:ddotf_range}) for $\tau\gtrsim1$\,kyr and $f_0\lesssim600$\,Hz and $\Delta \ddot{f_0} \sim 10^{-18}$\,Hz\,s$^{-2}$ from (\ref{eqn:para_metric}) and (\ref{eqn:bin_size}) with $T_{\rm drift}=10$\,d. In this example, only one value of $\ddot{f_0}$ is searched and the cost scaling in Equation (\ref{eqn:compute_time}) reduces to ${f_0}_{\rm max}^2 \tau^{-1} T_{\rm drift}^4$.\footnote{It is shown in Ref. \cite{Aasi2015-snr} that in the S6 search the computing cost scales approximately as ${f_0}_{\rm max}^{2.2}\tau^{-1.1}T_{\rm drift}^4$.} If we assume that only one value of $\ddot{f_0}$ is searched. For $\kappa=4\times 10^{-8}$\,s, $m=0.2$, $\beta = 1$, $N_{\rm ifo}=2$, $T_{\rm SFT}=1800$\,s and $N_{\rm core}=10^3$\,cores running in parallel, we obtain 
\begin{equation}
\label{eqn:compute_time_simplified}
\mathcal{T}_{f_0,\dot{f_0}} = 0.36\,{\rm d} \left(\frac{{f_0}_{\rm max}}{0.6\,{\rm kHz}}\right)^2\left(\frac{0.3\,{\rm kyr}}{\tau}\right)\left(\frac{T_{\rm drift}}{10\,{\rm d}}\right)^4\left(\frac{10^3}{N_{\rm core}}\right).
\end{equation}
However, for very young objects (e.g., $\tau \lesssim 0.03$\,kyr) with larger $\ddot{f_0}$, the cost scales as $\mathcal{T}_{f_0,\dot{f_0},\ddot{f_0}}\propto {f_0}_{\rm max}^3 \tau^{-3} T_{\rm drift}^7$.

\begin{figure}
	\centering
	\scalebox{0.36}{\includegraphics{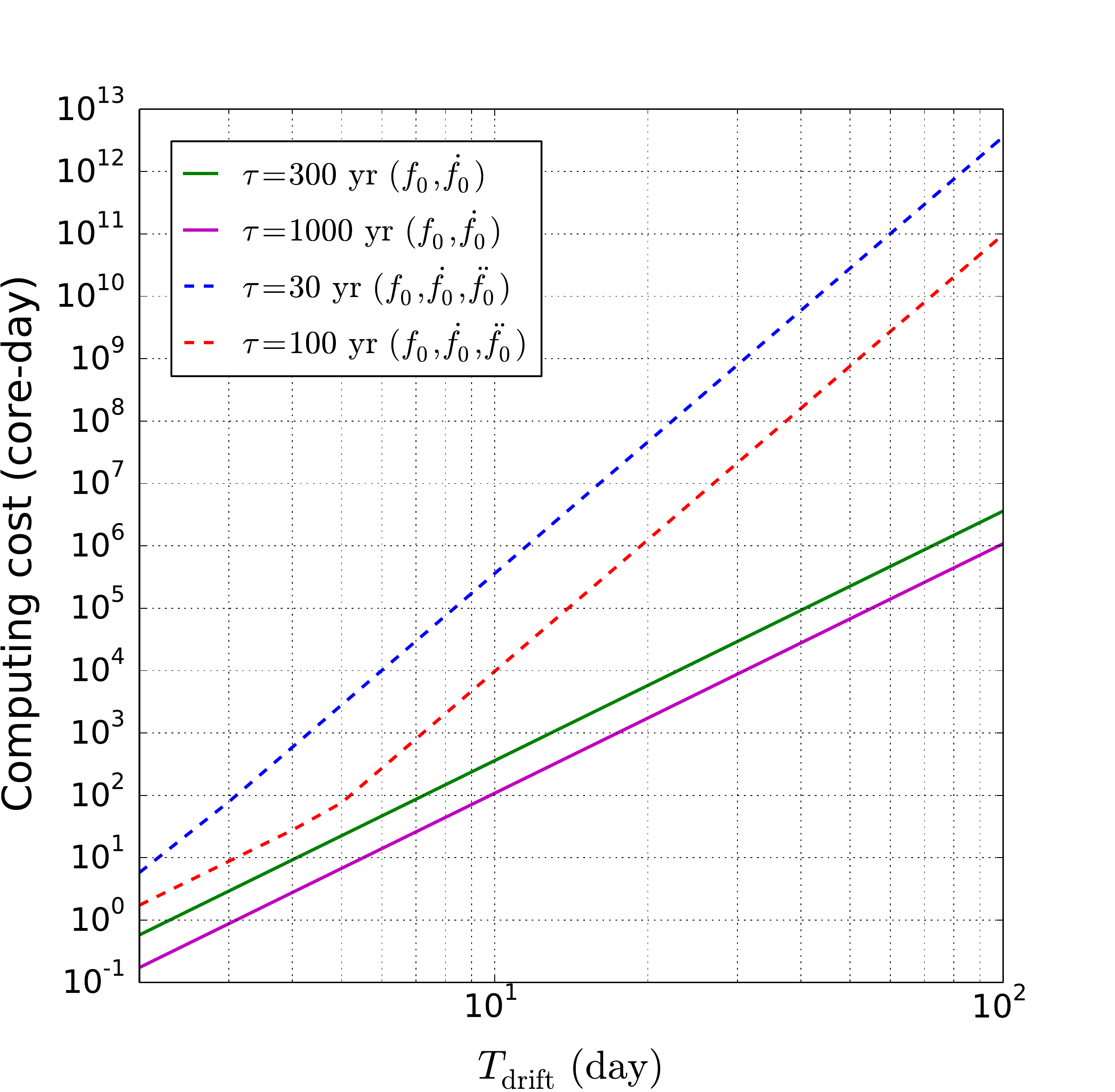}}
	\caption{Computing cost (in core-days) of a coherent $\mathcal{F}$-statistic search with $T_{\rm obs} = T_{\rm drift}$ as a function of $T_{\rm drift}$ (in days). The two solid curves (top and bottom) represent computing $\mathcal{F}(f_0, \dot{f_0})$ (or include a single $\ddot{f_0}$ value) for objects with $\tau = 0.3$\,kyr and 1\,kyr, respectively. The two dashed curves (top and bottom) represent computing $\mathcal{F}(f_0, \dot{f_0},\ddot{f_0})$ for objects with $\tau = 0.03$\,kyr and 0.1\,kyr, respectively. Parameters: ${f_0}_{\rm max}=600$\,Hz.}
	\label{fig:cost_ddotf}
\end{figure}

Figure \ref{fig:cost_ddotf} shows the cost of computing the $\mathcal{F}$-statistic over a coherent segment $T_{\rm drift}$ (in units of core-day). For concreteness, we fix ${f_0}_{\rm max} = 600$\,Hz. In a real search, ${f_0}_{\rm max}$ is a function of $\tau$, because we determine ${f_0}_{\rm max}$ to be the maximum frequency where the estimated strain sensitivity of the search beats the indirect spin-down limit [see Equation (\ref{eq:h0upperlim})]. If we compute $\mathcal{F}(f_0, \dot{f_0})$ (or search a single $\ddot{f_0}$ value), the costs for objects with $\tau = 0.3$\,kyr and 1\,kyr are indicated by the two solid curves. A coherent $\mathcal{F}$-statistic search or a stack-slide-based semi-coherent $\mathcal{F}$-statistic search requires searching higher-order derivatives for objects with $\tau \lesssim 0.3$\,kyr. The two dashed curves (top and bottom) represent the cost of computing $\mathcal{F}(f_0, \dot{f_0},\ddot{f_0})$ for objects with $\tau = 0.03$\,kyr and 0.1\,kyr, respectively. 

The serial clock time for computing can be reduced by parallelization. For $10^3$ nodes running in parallel, a coherent $\mathcal{F}$-statistic search over $T_{\rm drift}=10$\,d takes about 9\,hr for an object with $\tau = 0.3$\,kyr (e.g., Cas A), and about 10\,d for an object with $\tau = 0.1$\,kyr (e.g., G1.9+0.3). In reality, the cost indicated by the top dashed curve for an object with $\tau = 0.03$\,kyr (e.g., SNR 1987A) is still underestimated, because $\dddot{f_0}$ and higher-order derivatives must be searched using a stack-slide-based semi-coherent method.

\begin{figure}
	\centering
	\scalebox{0.37}{\includegraphics{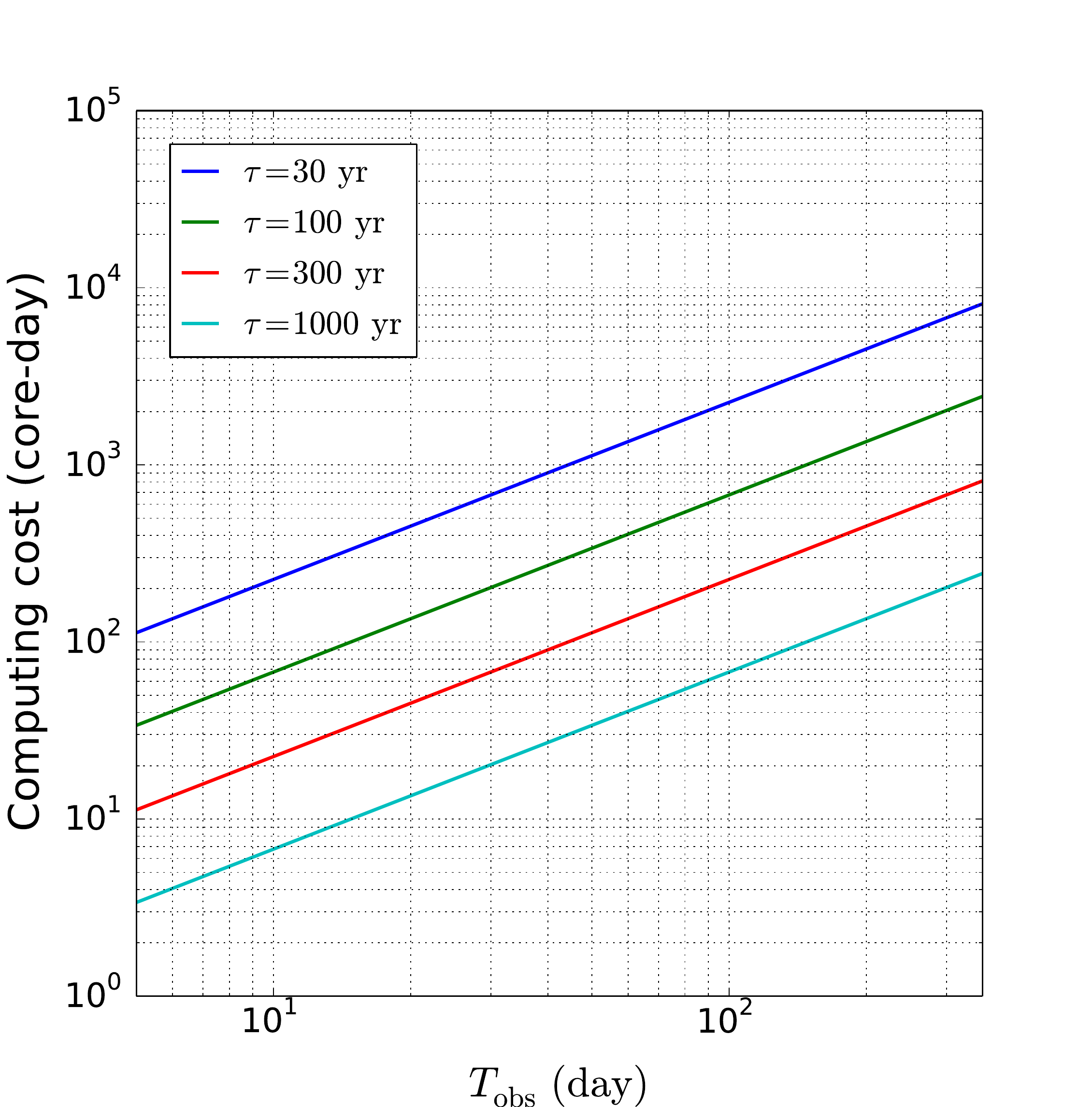}}
	\caption{Computing cost (in core-days) of a semi-coherent HMM search as a function of $T_{\rm obs}$ (in days) with $T_{\rm drift}$ fixed. The four curves (from top to bottom) represent objects with $\tau = 0.03$\,kyr, 0.1\,kyr, 0.3\,kyr and 1\,kyr, respectively. Parameters: ${f_0}_{\rm max}=600$\,Hz, $T_{\rm drift}=5$\,d, $N_T = T_{\rm obs}/T_{\rm drift}$.}
	\label{fig:cost_ddotf_hmm}
\end{figure}

We do not search second- or higher-order derivatives of the frequency in the HMM tracking. We start by discussing the ($f_0, \dot{f_0}$) tracking. In the two-dimensional tracking, we can always substitute the maximum $\ddot{f_0}$ in the range (\ref{eqn:ddotf_range}) into (\ref{eqn:int_T_drift_2}) to choose $T_{\rm drift}$ and ignore $\ddot{f_0}$ (or search a single value). However, for very young objects (e.g., $\tau \lesssim 0.03$\,kyr) with larger $\ddot{f_0}$, there can be sensitivity loss due to the short $T_{\rm drift}$ derived from (\ref{eqn:int_T_drift_2}). The relation between theoretical sensitivity and $T_{\rm drift}$ is discussed in Appendix \ref{sec:sensitivity}. The estimate of $\mathcal{T}$ in (\ref{eqn:compute_time_simplified}) stands for the time required to calculate the $\mathcal{F}$-statistic over one block of coherent segment $T_{\rm drift}$. For HMM tracking, we need to add up the time required to calculate all the required values of $\mathcal{F}(f_0,\dot{f_0})$. In addition there is a second component to the computing cost, namely solving the HMM. HMM tracking incoherently combines the $\mathcal{F}$-statistic outputs from $N_T=T_{\rm obs}/T_{\rm drift}$ blocks of data. The Viterbi algorithm computes $Q^*(O)$ via $(N_T+1)N_Q \ln N_Q$ operations \cite{Suvorova2016}. The total computing cost is dominated by the cost of computing $N_T$ blocks of the $\mathcal{F}$-statistic, scaling as $\mathcal{T} \propto N_T {f_0}_{\rm max}^2 \tau^{-1} T_{\rm drift}^4$. If we take $T_{\rm drift}=5$\,d as an example, Figure \ref{fig:cost_ddotf_hmm} shows the computing cost of a semi-coherent HMM search (in units of core-day) as a function of $T_{\rm obs}=N_T T_{\rm drift}$ for targets with $\tau = 0.03$\,kyr, 0.1\,kyr, 0.3\,kyr and 1\,kyr, respectively. In practice, $T_{\rm drift}>5$\,d is allowed for older targets and $T_{\rm drift}<5$\,d is required for younger ones.

When tracking $f_0$ alone, the search is always cheaper. We choose $\Delta f_0 = 1/(2 T_{\rm drift})$, satisfying $m \leq 0.2$. The metric given by (\ref{eqn:para_metric}) is no longer needed for $k=0$. Hence the number of templates and cost needed for computing $\mathcal{F}$-statistic over each block of $T_{\rm drift}$ in (\ref{eqn:num_templates}) and (\ref{eqn:compute_time}) reduce to
\begin{equation}
\label{eqn:num_templates_1d}
\mathcal{N} = 2 T_{\rm drift}{f_0}_{\rm max},
\end{equation}
and
\begin{equation}
\label{eqn:compute_time_1d}
\mathcal{T}_{f_0} = 2 \kappa \beta N_{\rm ifo}  T_{\rm drift}^2T_{\rm SFT}^{-1}{f_0}_{\rm max}.
\end{equation}
The total cost scales $\propto N_T {f_0}_{\rm max}T_{\rm drift}^2$ when tracking $f_0$ alone, saving a factor $\sim {f_0}_{\rm max}T_{\rm drift}^2$ compared to the ($f_0,\dot{f_0}$) tracking.

\section{$T_{\rm drift}$ given $\tau$}
\label{sec:estimate_T_drift}

We assume purely electromagnetic spin down ($\dot{f_0} \propto B_0^2f_0^n$, $n = 3$) for simplicity, which gives
\begin{equation}
\label{eqn:spin-down-eqn}
\dot{f_0}(t) = -\eta f_0(t)^3,
\end{equation}
where the coefficient $\eta \propto B_0^2$ is a positive constant. At time $t=0$ when the star was born, we have $f_0(t) = {f_0}_{\rm birth}$. The differential equation (\ref{eqn:spin-down-eqn}) has the solution
\begin{equation}
\label{eqn:soln-diff-eqn}
1- \left(\frac{f_0}{{f_0}_{\rm birth}}\right)^2 = - \frac{2 \tau \dot{f_0}}{f_0}.
\end{equation}
Substituting $\dot{f_0} = - (2T_{\rm drift}^2)^{-1}$ from (\ref{eqn:T_drift_form}) into (\ref{eqn:soln-diff-eqn}), we obtain
\begin{equation}
\label{eqn:soln-T-drift}
T_{\rm drift} = \tau^{1/2} f_0^{-1/2}\left[1-\left(\frac{f_0}{{f_0}_{\rm birth}}\right)^2\right]^{-1/2}.
\end{equation}

\section{Transition probability matrix for $f_0$, $\dot{f_0}$ tracking}
\label{sec:transition_matrix_proof}

We first derive the transition probabilities corresponding to the substate $\dot{f_0}$. Using Equations (\ref{eqn:dotf_range}) and (\ref{eqn:ddotf_range}), the range of $\ddot{f_0}$ is given by
\begin{equation}
\frac{0.2{f_0}_{\rm min}}{\tau^2}\lesssim \ddot{f_0}\lesssim \frac{2{f_0}_{\rm max}}{\tau^2},
\end{equation}
where ${f_0}_{\rm min}$ and ${f_0}_{\rm max}$ are the minimum and maximum $f_0$ being searched. The maximum $\ddot{f_0}$ is more than two orders of magnitude larger than the minimum $\ddot{f_0}$. We assume that $\ddot{f_0}$ is uniformly distributed in the range $0\leq \ddot{f_0}\leq \ddot{f_0}_{\rm max}$. At each step, given Equation (\ref{eqn:int_T_drift_2}), $\dot{f_0}$ jumps at most one bin up or stays in the same bin with equal probability 1/2.

We then estimate the number of bins $f_0$ moves during each step. Equation (\ref{eqn:next_f0}) is more precisely given by
\begin{eqnarray}
\nonumber f_0(t_{\rm n}) + \dot{f_0} (t_{\rm n})T_{\rm drift}  
\leq f_0(t_{\rm n+1}) \leq  \\
f_0(t_{\rm n}) + [\dot{f_0}(t_{\rm n})+\Delta \dot{f_0}] T_{\rm drift}.
\end{eqnarray}
Let us write $q(t_{\rm n})=[f_0(t_{\rm n}),\dot{f_0}(t_{\rm n})]=q_{ij}$, where $i$ and $j$ index $f_0$ and $\dot{f_0}$ bins, respectively. Then the number of bins that $f_0$ moves from step $t_{\rm n}$ to step $t_{\rm n+1}$, denoted by $\Delta i$, takes the minimum and maximum values 
\begin{eqnarray}
\Delta i_{\rm min} (j)&=& { \rm floor}(|\dot{f_0}_{j+1}|T_{\rm drift}/\Delta f_0), \\
\Delta i_{\rm max} (j)&=& {\rm ceil}(|\dot{f_0}_j|T_{\rm drift}/\Delta f_0),
\end{eqnarray}
where ${\rm floor}(x)$ denotes the largest integer smaller than or equal to $x$, ${\rm ceil}(x)$ denotes the smallest integer larger than or equal to $x$, and $\dot{f_0}_j$ is the value of $\dot{f_0}$ in the $j$-th $\dot{f_0}$ bin. In other words, $f_0(t_{\rm n+1})$ can be located in any bin within the range $[i-\Delta i_{\rm max}, i-\Delta i_{\rm min}]$ with uniform probability.\footnote{Since $\dot{f_0}$ is negative, we always have $f_0(t_{\rm n+1})\leq f_0(t_{\rm n})$.} The two dimensional transition probability matrix is given by
\begin{eqnarray}
A_{q_{i-\Delta i,  j+1} q_{ij}} &=& A_{q_{i-\Delta i,  j} q_{ij}} \\
&=& \{2[\Delta i_{\rm max}(j) - \Delta i_{\rm min}(j)+1]\}^{-1},
\end{eqnarray}
where $\Delta i$ takes integer values $\Delta i_{\rm min}(j) \leq \Delta i \leq \Delta i_{\rm max}(j)$, and all other terms are zero.

\section{Analytic sensitivity scalings}
\label{sec:sensitivity}
In this section, we present an approximate analytic formula for the search sensitivity, based on a few general assumptions. Deviations are discussed in detail in Ref. \cite{Wette2012,Prix2012}. Accurate sensitivity scalings require Monte-Carlo simulations for each implementation of the search, as shown in Section \ref{sec:simulations_and_sensi}. 

The sensitivity of a search can be defined in terms of the characteristic gravitational-wave strain corresponding to 95\% detection efficiency. For a coherent $\mathcal{F}$-statistic search over one block of $T_{\rm drift}$, searching up to the highest frequency derivative required for a given mismatch, it takes the form \cite{Wette2008,Aasi2015-snr}
\begin{equation}
\label{eqn:sensi-95}
h_0^{95\%}(f)=\Theta S_h(f)^{1/2}(\beta T_{\rm drift})^{-1/2},
\end{equation}
where $\Theta$ is a statistical threshold, depending on the shape of the parameter space manifold. One finds $30\lesssim \Theta \lesssim 40$ for a directed search of the type discussed in this paper \cite{Wette2008}. The term $\beta T_{\rm drift}$ gives the length of the interferometer data in the timespan $T_{\rm drift}$.

As every block of $\mathcal{F}$-statistic output over $T_{\rm drift}$ is chi-squared distributed with four degrees of freedom,\footnote{Here we assume that the $\mathcal{F}$-statistic is independently and identically distributed. The estimate requires modification when applied to real interferometer data, where the noise is non-stationary and/or non-Gaussian. A more robust Bayesian framework is introduced in Ref. \cite{Keitel2014} to analyze the $\mathcal{F}$-statistic in the presence of instrumental artifacts.} and the chi-squared distribution is additive, we can calculate the PDF of $z=\ln P(Q|O)$ along the true signal path from (\ref{eqn:chi2-dist}) and (\ref{eqn:snr2}) by multiplying both the degrees of freedom and the noncentrality parameter by $N_T=T_{\rm obs}/T_{\rm drift}$. If $Q^*(O)$ coincides exactly with the true path, we obtain
\begin{equation}
\label{eqn:N-chi2-dist}
p(z) = \chi^2\left[z; \frac{4T_{\rm obs}}{T_{\rm drift}},\frac{K h_0^2T_\text{obs}}{S_h(f)}\right],
\end{equation}
If $Q^*(O)$ does not intersect the true path anywhere, we have
\begin{equation}
\label{eqn:N-chi2-dist-noise}
p(z) = \chi^2\left(z; \frac{4T_{\rm obs}}{T_{\rm drift}},0\right),
\end{equation}
Combining (\ref{eqn:N-chi2-dist}) and (\ref{eqn:N-chi2-dist-noise}), the signal-to-noise ratio after $N_T$ steps of the HMM equals $\rho_0'$, given by
\begin{eqnarray}
\rho_0'^2 &=& \frac{\mu_{\rm signal} -\mu_{\rm noise}}{\sigma_{\rm noise}}\\
&\propto&\frac{h_0^2}{S_h(f)}(T_{\rm obs}T_{\rm drift})^{1/2},
\end{eqnarray}
where $\mu_{\rm signal}=N_T\rho_0^2$ and $\mu_{\rm noise}=0$ are the noncentralities of the distributions in (\ref{eqn:N-chi2-dist}) and (\ref{eqn:N-chi2-dist-noise}), respectively, and $\sigma_{\rm noise}=(8N_T)^{1/2}$ is the standard deviation of the distribution in (\ref{eqn:N-chi2-dist-noise}). Hence we obtain
\begin{equation}
\label{eqn:sensi-scaling}
h_0^{95\%}(f) = \Theta S_h(f)^{1/2} (T_{\rm obs}T_{\rm drift})^{-1/4},
\end{equation}
assuming $\beta = 1$. When $T_{\rm obs} = T_{\rm drift}$, equation (\ref{eqn:sensi-scaling}) reduces to (\ref{eqn:sensi-95}).

\begin{figure*}
	\centering
	\subfigure[]
	{
		\label{fig:h0_vs_f}
		\scalebox{0.265}{\includegraphics{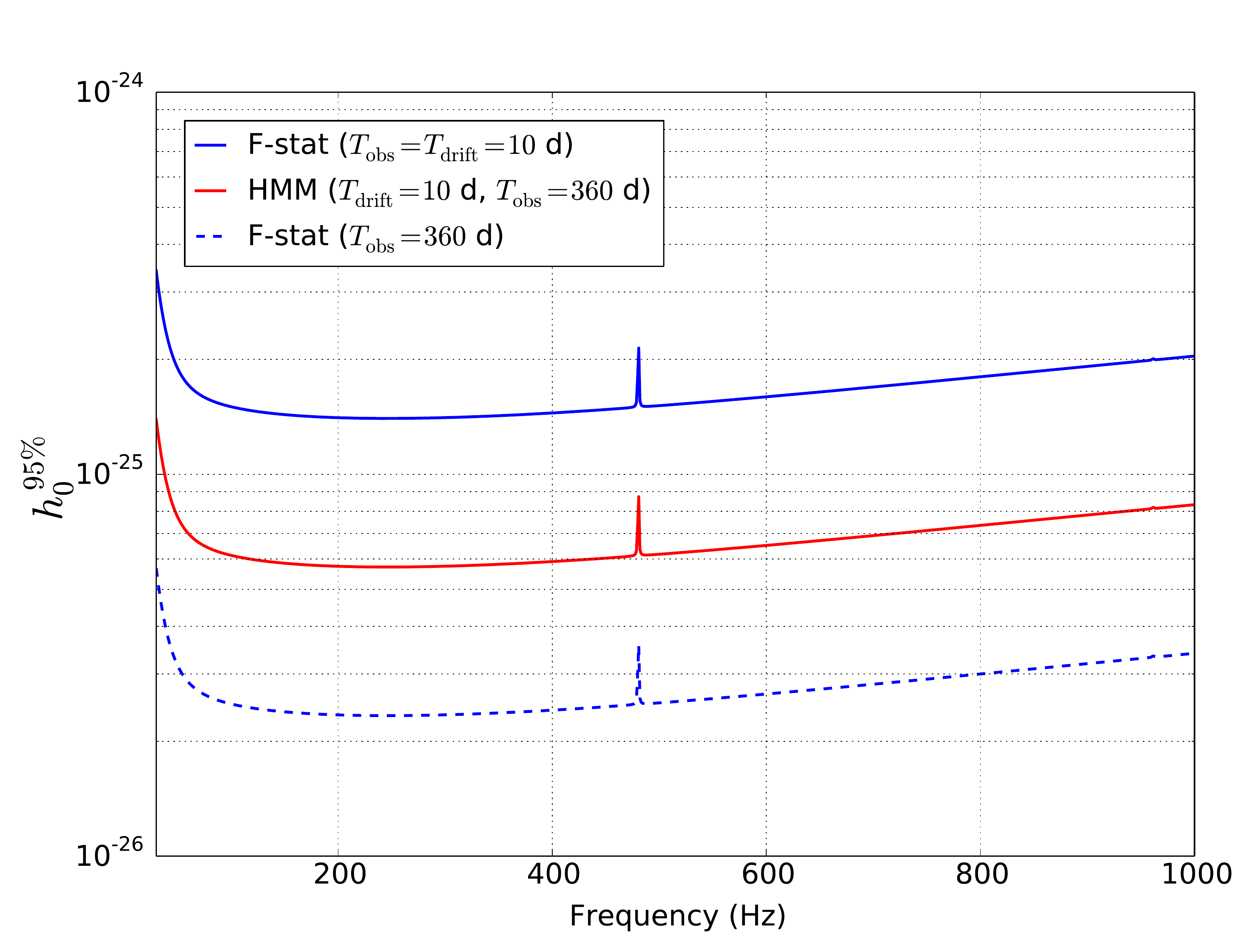}}
	}
	\subfigure[]
	{
		\label{fig:h0_vs_T}
		\scalebox{0.26}{\includegraphics{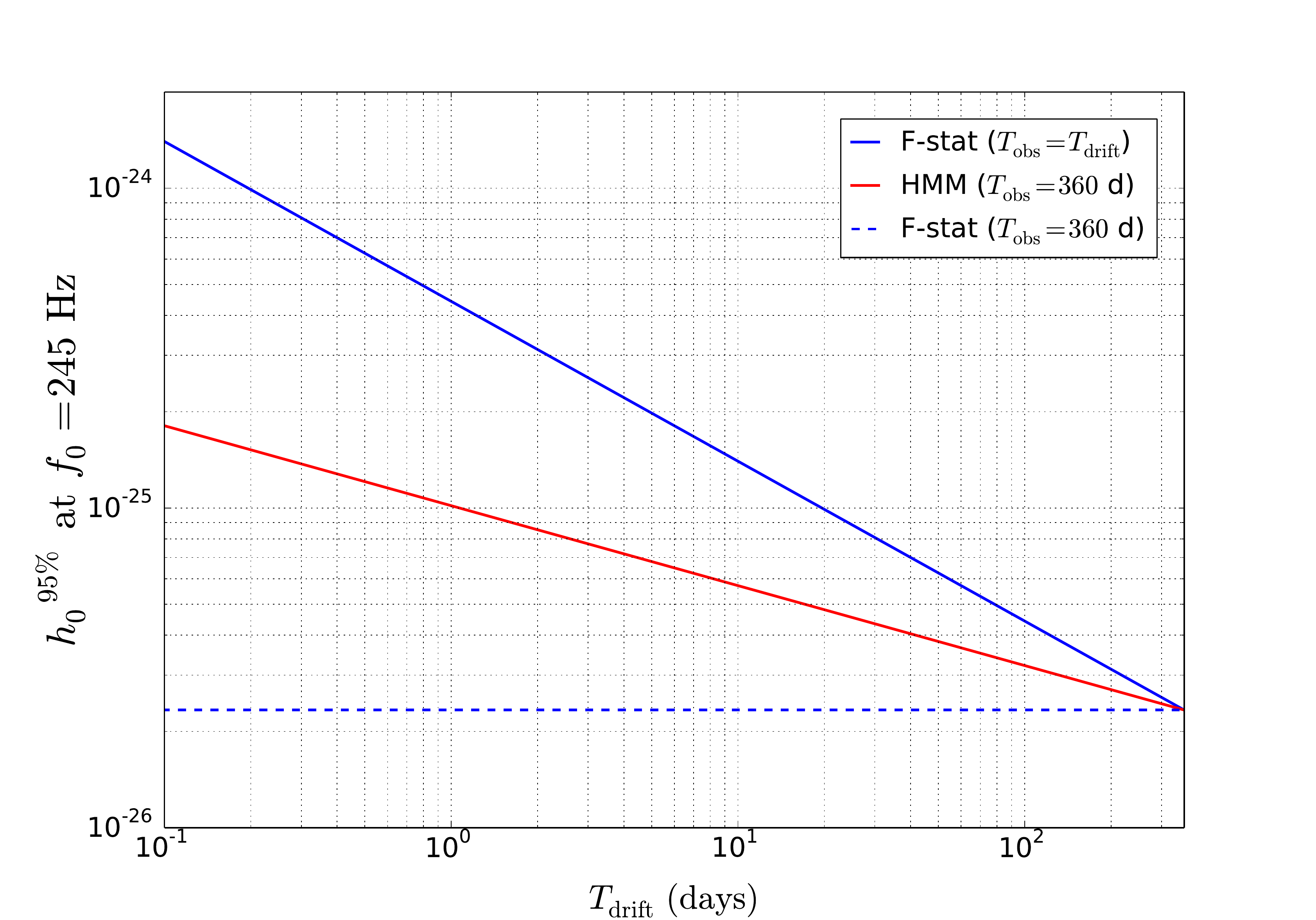}}
	}
	\caption{Analytical sensitivity $h_0^{95\%}$ of two coherent $\mathcal{F}$-statistic searches (blue curves) and a semi-coherent HMM search (red curve) for an isolated young neutron star at the design sensitivity of Advanced LIGO. The blue solid curve indicates a coherent $\mathcal{F}$-statistic search of duration $T_{\rm drift}$, the maximum practical duration set by spin down in Ref. \cite{Aasi2015-snr}. The blue dashed curve indicates a coherent $\mathcal{F}$-statistic search with $T_{\rm obs}=360$\,d. The red curve indicates a HMM search with $T_{\rm obs}=360$\,d and $N_T = T_{\rm obs}/T_{\rm drift}$. (a) $h_0^{95\%}$ as a function of signal frequency $f_0$. (b) Minimum $h_0^{95\%}$ [i.e., $h_0^{95\%}$ achieved at 245\,Hz in (a)] as a function of $T_{\rm drift}$. The HMM sensitivity scales as $\sim T_{\rm drift}^{-1/4}$ for fixed $T_{\rm obs}$. When $T_{\rm drift}$ is fixed, the HMM (red curve) improves upon the $\mathcal{F}$-statistic search (blue solid curve) by a factor of $\sim N_T^{1/4}$. The spikes in the first panel are instrumental lines predicted to occur in Advanced LIGO at its design sensitivity. (Statistical threshold: $\Theta=35$)}
	\label{fig:sensitivity}
\end{figure*}

By way of illustration, we compute $h_0^{95\%}$ for HMM tracking with $T_{\rm obs}=N_T T_{\rm drift} = 360$\,d and compare the result with a coherent $\mathcal{F}$-statistic search limited to $T_{\rm obs} = T_{\rm drift}$ like in Ref. \cite{Aasi2015-snr}. We take $\Theta = 35$ and let $S_h(f)$ be the Advanced LIGO design noise PSD in (\ref{eqn:sensi-95}). Figure \ref{fig:sensitivity} shows the results. The blue solid curve, red solid curve, and blue dashed curve indicate an $\mathcal{F}$-statistic search of duration $T_{\rm drift}$, a HMM search of duration $N_T T_{\rm drift}$, and a fully coherent $\mathcal{F}$-statistic search with $T_{\rm obs}=360$\,d, respectively. Figure \ref{fig:h0_vs_f} plots $h_0^{95\%}$ as a function of signal frequency. When $T_{\rm drift}$ is fixed [e.g., $T_{\rm drift}=10$\,d in Figure \ref{fig:h0_vs_f}], the HMM tracking of duration $N_T T_{\rm drift}$ improves upon the sensitivity of the $\mathcal{F}$-statistic search of duration $T_{\rm drift}$ by a factor of $\sim N_T^{1/4}$. Figure \ref{fig:h0_vs_T} plots the minimum $h_0^{95\%}$ in Figure \ref{fig:h0_vs_f} achieved in the band where the detectors are most sensitive (at 245\,Hz) as a function of $T_{\rm drift}$. The sensitivity achievable by the HMM scales as $\sim T_{\rm drift}^{-1/4}$ for fixed $T_{\rm obs}=1$\,yr. Figures \ref{fig:h0_vs_f} and \ref{fig:h0_vs_T} demonstrate together the scaling indicated by (\ref{eqn:sensi-scaling}). A fully coherent search using all the data (duration $N_T T_{\rm drift}$) indicated by the blue dashed curve is more sensitive than the HMM of course. However it is computationally expensive.\footnote{One fully coherent search directed at the isolated compact object Calvera has been carried out, using Initial LIGO S5 data ($T_{\rm obs}=2$\,yr). It is based on a resampling technique and saves computing cost by a factor $T_{\rm obs}(\log T_{\rm obs})^{-1}$, yielding a minimum upper limit $h_0 ^{95\%}=1.14\times 10^{-25}$ at 152\,Hz \cite{Patel2011}.} The theoretical scalings here also apply approximately to other $\mathcal{F}$-statistic-based semi-coherent searches.

\end{document}